\documentclass{aa}
\usepackage{graphicx}
\usepackage{txfonts}
\usepackage{amsmath}
\usepackage{amsfonts}
\usepackage{amssymb}
\usepackage{url}
\usepackage{subfig}
\usepackage{tabularx}
\usepackage{natbib}
\bibpunct{(}{)}{;}{a}{}{,} 
\newcommand\T{\rule{0pt}{2.6ex}}       

\begin{document}

   \title{The SuperWASP catalogue of 4963 RR Lyr stars: identification of 983 Blazhko effect candidates.}

   \titlerunning{The SuperWASP catalogue of RR Lyr stars.}

   \author{P. A. Greer\inst{1} \and S. G. Payne\inst{1} \and A.J. Norton\inst{1} \and P. F. L. Maxted\inst{2} \and B. Smalley\inst{2} \and R.G. West\inst{3} \and P. J. Wheatley\inst{3}  \and U. C. Kolb\inst{1} }

    \authorrunning{Greer et al.}

   \institute{School of Physical Sciences, The Open University, Walton Hall,
   Milton Keynes MK7 6AA, U.K.
   \and
   Astrophysics Group, Keele University, Staffordshire ST5 5BG, U.K.
   \and
   Department of Physics, University of Warwick, Coventry CV4 7AL, U.K.
   }

   \date{Received December 20, 2017; accepted July 3, 2017}


  \abstract
 {}
   {We set out to compile a catalogue of RRab pulsating variables in the SuperWASP archive and identify candidate Blazhko effect objects within this catalogue. We analysed their light curves and power spectra for correlations in their common characteristics to further our understanding of the phenomenon.}
   {Pulsation periods were found for each SWASP RRab object using PDM techniques. Low frequency periodic signals detected in the {\sc CLEAN} power spectra of RRab stars were matched with modulation sidebands and combined with pairs of sidebands to produce a list of candidate Blazhko periods. A novel technique was used in an attempt to identify Blazhko effect stars by comparing scatter at different parts of the folded light curve. Pulsation amplitudes were calculated based on phase folded light curves.}
   {The SuperWASP RRab catalogue consists of 4963 objects of which 3397 are previously unknown. We discovered 983 distinct candidates for Blazhko effect objects, 613 of these being previously unknown in the literature as RR Lyrae stars, and 894 are previously unknown to be Blazhko effect stars. Correlations were investigated between the scatter of points on the light curve, the periods and amplitudes of the objects' pulsations, and those of the Blazhko effect.
   }
   {A statistical analysis has been performed on a large population of Blazhko effect stars from the wide-field SuperWASP survey. No correlations were found between the Blazhko period and other parameters including the Blazhko amplitude, although we confirmed a lower rate of occurrence of the Blazhko effect in long pulsation period objects. }


   \maketitle
%

\section{Introduction}
RR Lyrae stars (RRL) are pulsating variable stars with pulsation periods ($P_{\rm{pulse}}$) around 0.5~d. They are horizontal branch (HB), helium burning stars, therefore mainly belonging to Population II with some old Population I. Their distinctive pulsations and limited range of intrinsic brightness make them both excellent distance markers \citep{ngeow_distance_2013} and age identifiers \citep{fiorentino_ancient_2012} for globular clusters.

Stellar pulsations in the mid instability strip are understood to be caused by the $\kappa$ mechanism \citep{percy_pulsating_2009} where ionisation zones decouple pressure and temperature, trapping heat until the ionisation layer is pushed out past the point of equilibrium thereby allowing the "heat-piston" like effect to become cyclic. RR Lyrae stars are classified using Bailey classes \citep{bailey_discussion_1902}, according to the shape of their light curves as RRab, RRc or RRd, which depend on whether they pulsate in the fundamental mode, first overtone, or both, respectively. The General Catalogue of Variable Stars (GCVS; \citealp{samus_vizier_2009}) provides a list of over 6,000 RRL along with their sub-classes.
\cite{szabo_automated_2004} demonstrated how these modes are not clearly separated, and with certain metallicities and masses the mode depends on whether the star is evolving redwards or bluewards through the instability strip.

The Blazhko effect, named after Sergei Blazhko who discovered it in 1907 \citep{blazko_mitteilung_1907}, consists of both amplitude modulation (AM) and frequency modulation (FM) of the pulsations of RR Lyrae. The effect is generally periodic but recent exceptions have been discovered with varying Blazhko periods ($\rm{P_{BL}}$) \citep{le_borgne_stellar_2007}. The Blazhko effect of the prototype RR Lyrae itself appears to be variable \citep{kolenberg_blazhko_2006}, have an extended 14 year $\rm{P_{BL}}$ \citep{le_borgne_historical_2014}, and stopped for 4 years \citep{jurcsik_modulation_2014}. Space based observations by \textit{Kepler} have led to the discovery of the interlacing of two pulsation periods with double the originally observed pulsation periods called "period doubling" \citep{szabo_does_2010}. This can be modelled by a 9:2 resonance between the fundamental pulsation mode and the surface 9th overtone mode \citep{buchler_blazhko_2011}. Since RRLyr have periods of about half a day, only alternate pulsations can be observed from a single ground-based site, making this effect extremely difficult to observe from the ground. Radial-non-radial pulsation mode combinations have also been identified in double--mode RR Lyrae (RRd) stars in the OGLE-IV campaign \citep{netzel_double-mode_2015}.

Space based observations such as those by the \textit{CoRoT} and \textit{Kepler} satellites have produced detailed studies, with the \textit{Kepler} mission being extended into the \textit{K2} campaign \citep{howell_k2_2014,molnar_rr_2015}. Large catalogues of RR Lyrae have been created from ground-based observations including the Optical Gravitational Lensing Experiment (OGLE) ground based survey \citep{soszynski_over_2014} which has recorded 38,257 RR Lyr objects in the Galactic bulge, including RRd exhibiting the Blazhko effect.
RR Lyrae have also been catalogued by The Catalina Transient Real--Time Survey \citep{drake_first_2009} and LINEAR \citep{stokes_lincoln_2000} programs.
The MACHO survey \citep{alcock_macho_2003} identified many of the pulsation characteristics from a large sample of over six thousand stars from the Large Magellanic Cloud, but this collection is known to have a lower metallicity than halo field or cluster stars. \citet{skarka_blasgalf_2013} has produced a web-based compilation of over 400 Blazhko stars, but, as stated on the web site, it does not include objects from densely populated areas such as globular clusters or the Galactic bulge, in much the same way as WASP (see below), nor objects from the OGLE, LINEAR or Catalina surveys mentioned above. However, Skarka has investigated bright RRab objects from the WASP dataset \citep{skarka_bright_2014} in collaboration with data from ASAS (All Sky Automated Survey; \citealp{pojmanski_all_1997}).
\cite{jurcsik_konkoly_2009} suggested that the proportion of RRL showing the Blazhko effect approaches 50\%, with the review by \cite{kovacs_blazhko_2016} mentioning how the occurrence rate in ground--based observations do not lag far behind space--based surveys. It is therefore an important feature of pulsating stars, not present in just a few isolated cases. This higher proportion has led to theories that it is a common but usually undiscovered feature of most RRL, despite different metallicities and dominant pulsations modes.
Unlocking the secrets of the Blazhko effect will therefore greatly improve our understanding of stellar pulsations.

Despite this interest, the physical cause of the Blazhko effect is unknown. Current models to explain the Blazhko effect include non-resonant beating-modes \citep{bryant_is_2015}, shockwaves originating in the stellar atmosphere \citep{gillet_atmospheric_2013} and resonant modes \citep{szabo_revisiting_2014} as mentioned above (see \cite{kovacs_blazhko_2016} for a critical comparison of current models). There are a wide range of both pulsation and Blazhko effect periods and amplitudes; the Blazhko effect does not appear to be tied to a single property of the star or its pulsation. An alternative approach is to define characteristic attributes and then look for correlations between those attributes and features in their distributions. Investigations like this benefit from a large, spatially well distributed population of relatively bright stars; such a population is available in the all-sky SuperWASP survey.

WASP is the Wide Angle Search for Planets project \citep{pollacco_wasp_2006} based at installations in La Palma in the Canary Islands and Sutherland in South Africa, providing an almost all sky survey since 2006. Each SuperWASP (SWASP) instrument consists of 8 Canon 200 mm f/1.8 lenses each with CCDs that have an angular scale of 13.7~arcseconds per pixel, mounted on a Torus mount allowing a cadence of approximately 10~minutes. It uses wide optical bandwidth filters from 400 to 700 nm to record objects down to 15th magnitude. Aperture photometry from the 3.5 pixel (34'' radius) aperture \citep{norton_new_2007} is detrended using the SysRem algorithm \citep{tamuz_correcting_2005} producing the flux measurements used throughout this work. Objects are identified using the USNO-B1.0 input catalogue and given a unique 1SWASP identifier (SWASPid) based on these coordinates. The archive contains over 31 million objects.

The aims of this research are to further our knowledge of the Blazhko effect by cataloguing the photometric pulsation characteristics of a large sample of Blazhko stars harvested from the SWASP archive. Blazhko effect stars are identified and the frequencies and amplitudes of a large range of Blazhko effects are collated. Relations between the main pulsation and those of the Blazhko effect are then identified and statistically analysed.
This will allow detailed follow up of individual objects, such as more in--depth chemical and metallicity analysis, or further frequency domain findings. Results can also be tested against the expected attributes from theoretical models.

The rest of this paper is set out in the following sections: Sec.~\ref{method} includes subsections describing the SWASP data, phase folding of light curves and comparing SWASP objects to other catalogues; Sec.~\ref{sec:FreqDomain} explains how Fourier spectra produced by the {\sc CLEAN} algorithm were used to look for the Blazhko effect in the frequency domain; Sec.~\ref{sec:TimeDomain} describes investigations of the Blazhko effect in the time domain; Sec.~\ref{sec:analysis} describes the processes of looking for correlations between the parameters; the paper finishes with Discussion (Sec.~\ref{sec:discussion}) and Conclusions (Sec.~\ref{sec:conclusions}).

\section{Method}
\label{method}
\subsection{SuperWASP data}
\label{SuperWASPdata}
A machine learning algorithm by \cite{payne_identification_2013} made a preliminary identification and classification of variable objects in the SWASP archive based on the shape of their phase folded light curves. The algorithm was capable of identifying sub--categories of pulsating variable objects to the level of segregating RR Lyrae objects into their standard subclasses of RRab and RRc. It produced a list of the SWASP unique identifiers and pulsation periods for 8556 potential RRab objects. These objects comprised our initial sample and were extracted from the SWASP archive. Cameras 281 -- 288, operational at SuperWASP South from July 2012, use different filters and lenses to capture brighter objects \citep{turner_finding_2015}. Data from these cameras were recorded at a higher cadence but lower signal-to-noise ratio and were removed to maintain a quality consistent with SuperWASP North objects.
Outlier data were also removed at this stage. Firstly, spurious data points containing negative fluxes were removed. Next, an iterative clipping algorithm based on \cite{holdsworth_high-frequency_2014} was used, where flux outliers at greater than $5\sigma$ from the mean were removed and the remaining data were fed back into the clipping routine to be clipped at $5\sigma$ again, for a total of 5 iterations.
Together, this removed a median of 0.18\% points and a mean of 1.5\% points. The typical light curve used throughout this work consists of ~30,000 data points with an average duration of 5.5 years, offering the opportunity to directly detect Blazhko effect signals with frequencies of up to a year.

\subsection{Phase folding of light curves}
\label{sec:PhaseFoldedLC}
The initial 8556 RRab objects were checked by inspecting their light curves. The phase folding was performed using a custom phase dispersion minimisation (PDM) and epoch folding routine based on \cite{davies_improved_1990}, where the aim is to minimise the $\chi^2$ value within each bin while maximising the $\chi^2$ value from the bin medians across the entire epoch. The initial parameters were set to use 50~bins with a range of trial periods centred on the period given by \cite{payne_identification_2013}.
To find the pulsation period for each object the folding routine used 3 iterations, decreasing the width of the trial period range from 2000~s to 10~s, then to 1~s, while concurrently reducing the interval steps from 2~s, to 0.01~s, to 0.001~s so as to maintain resolution. The epoch folding scores were calculated as
\begin{equation}
  \chi^2_{EF} = \sum_{bin=1}^{M}{\frac{\left(\tilde{F}_{\rm{bin}} - \tilde{F}_{\rm{LC}}\right)^2}{\tilde{F}_{\rm{LC}}}} \nonumber
	\label{eq:EF}
\end{equation}
where $\tilde{F}_{\rm{bin}}$ is the median flux in each bin, $M$ is the number of bins and $\tilde{F}_{\rm{LC}}$ is the  median flux of the whole light curve.

The PDM reduced $\chi^2$ value for each bin was calculated as
\begin{equation}
  \chi^2_{red} = \sum_{j=1}^{M}{\left(\frac{ \Sigma_{i=1}^{i=N_{bin}}{\left(F_{i} - \tilde{F}_{\rm{bin}}\right)^2} }{\tilde{F}_{\rm{bin}}} \right) / \left(N_{bin}-1\right)} \nonumber
	\label{eq:PDM}
\end{equation}
where $F_{i}$ is the $i$th flux point in the bin, $\tilde{F}_{\rm{bin}}$ is the median flux in the bin, and $N_{bin}$ is the number of flux points in the bin, and $M$ is the number of bins. The best period was the trial period with the highest ratio of epoch folding to PDM values.
Visual inspections led to 2876 SuperWASP objects (33.6\%) being removed: either the re-folded light curve was flat with a gap due to a periodic gap in observations; or the light curves looked like a non-variable star being contaminated by light from a nearby bright pulsating variable, identifiable by the similarity in their light curve shapes and coordinates.
A further visual check was made on objects with periods below 0.32 d. If their light curve looked similar to the RRc type, their classification was checked using the online Vizier system. A further 20 objects were removed at this stage due to being classified as RRc in AAVSO or GCVS. Objects with periods above 1.2~d were also checked online and a further 9 were removed as they were classified as BL Her type objects.
It was now possible to cross reference this remaining dataset of 5651 RRab against known RRab and known Blazhko effect stars (which resulted in the removal of more duplicate objects as described below). The phase folded light curves could also be analysed for signs of the Blazhko effect, see Sec.~\ref{sec:TimeDomain}.

\subsection{Catalogue comparisons}
\label{CatalogueComparisons}
The remaining SWASP RRab objects were compared to 8076 'RR', 'RR:', 'RRAB' and 'RRAB:' objects in the General Catalogue of Variable Stars (GCVS) \citep{samus_vizier_2009}. The 'RRAB:' and 'RR:' lists were included in order to avoid declaring any existing objects as new variable stars. The SWASP RRab coordinates were taken directly from their unique SWASP identifier.
This cone search was repeated with 404 known Blazhko effect stars from M. Skarka's BlaSGalF \footnote{http://www.physics.muni.cz/~blasgalf/} website \citep{skarka_blasgalf_2013}. These known Blazhko effect stars include several RRc sub-type objects which could overlap with GCVS 'RR' or 'RR:' stars, or a SuperWASP star if the light curve could have been mistaken for an RRab, so they were retained during the catalogue comparison stage. Objects were declared either as known RRab objects, known Blazhko stars, or both if they were within 2 arcminutes of objects in the GCVS and~/~or Skarka's list.
This radius was selected after correctly matching a SWASP RRab to a known Blazhko star at 100 arcseconds while a larger radii risked falsely matching objects.

The comparison algorithm had the additional benefit of highlighting when several SuperWASP objects fell within the same search aperture. Some of these objects would be the same object given different coordinates when compared to the USNO-B1 catalogue, or a non-variable object misclassified as RRab through being contaminated by variable flux overspilling from a nearby bright RRab. The closest SuperWASP object to the GCVS or Skarka object was kept as the matching one and remaining objects within the 2 arcminute radius were listed as duplicates. To identify duplicate RRab within the SWASP RRab dataset the cone search routine was run again with the remaining SuperWASP objects being used as both input catalogues. Duplicates were defined as those stars within the same search radius, with periods within 1.5~s of each other. After manually checking light curves of several groups of duplicate objects, the object with the highest amplitude was selected to remain as the distinct RRab.
4963 RRab objects remained once 688 duplicates were removed. A sample of the SWASP RRab catalogue is shown in Table~\ref{tab:RRabcat}.

These 4963 RRab objects were further compared with CRTS data, finding 1,692 RRab matches in \cite{torrealba_discovery_2015} and 190 in \cite{drake_catalina_2014}. The Siding Springs Survey (SSS) \citep{torrealba_discovery_2015} identifier has been added to the SWASP RRab catalogue where CDS returned a match. In cases where there was no SSS match but there was a Catalina Schmidt Survey (CSS) \citep{drake_catalina_2014} match, the CSS identifier has been used.

\begin{table*}
	\caption{Subset of the SWASP RRab catalogue}
	\label{tab:RRabcat}
\begin{tabular}{l l l l l l} \hline \hline
 SWASPid & $P_{\rm{Pulse}}\ $ [d] & $A_{\rm{LC}}$ [mag] & $\bar{m}_{\rm{LC}}$ [mag] & GCVS name & SSS / CRTS name \T  \\
 \hline \T
1SWASPJ000003.66+352146.1 & 0.70675755 & 0.8$\pm$0.2 & 13.2$\pm$0.2 & GM And  &  \\ 1SWASPJ000017.72-101317.1 & 1.14682543 & 0.9$\pm$0.7 & 14.6$\pm$0.5 &  &  \\ 1SWASPJ000018.15+193255.3 & 0.54549623 & 0.8$\pm$1.0 & 15.9$\pm$0.7 & V0420 Peg  &  \\ 1SWASPJ000023.75+361942.7 & 0.56246346 & 0.8$\pm$0.5 & 14.6$\pm$0.4 &  &  \\ 1SWASPJ000035.58+263949.6 & 0.56606638 & 0.8$\pm$0.4 & 13.5$\pm$0.3 & GV Peg  &  \\ 1SWASPJ000049.23-284900.4 & 0.58303469 & 0.5$\pm$0.6 & 14.8$\pm$0.4 & TZ Scl  &  \\ 1SWASPJ000126.73-001344.2 & 0.55574918 & 0.5$\pm$0.6 & 14.9$\pm$0.4 &  &  \\ 1SWASPJ000128.97-411541.5 & 0.50587386 & 1.3$\pm$1.0 & 15.5$\pm$0.7 &  & J000129.0-411542 \\ 1SWASPJ000152.61+473311.6 & 0.71633267 & 0.6$\pm$0.5 & 14.4$\pm$0.3 &  &  \\ 1SWASPJ000157.75-364042.4 & 0.63752764 & 0.5$\pm$0.4 & 13.8$\pm$0.2 & CY Scl  & J000157.8-364043 \\ 1SWASPJ000216.95+452846.0 & 0.55842459 & 0.2$\pm$0.5 & 14.3$\pm$0.4 &  &  \\ 1SWASPJ000248.10-245643.1 & 0.49335495 & 1.1$\pm$0.4 & 10.4$\pm$0.2 & RU Scl  &  \\ 1SWASPJ000321.17+032352.4 & 0.57904023 & 0.8$\pm$0.4 & 13.8$\pm$0.3 &  &  \\ 1SWASPJ000349.42+370415.7 & 0.69132727 & 0.3$\pm$0.2 & 13.0$\pm$0.1 &  &  \\ 1SWASPJ000400.87-424356.7 & 0.51894063 & 0.6$\pm$0.5 & 14.7$\pm$0.3 & DK Phe  & J000400.8-424357 \\ 1SWASPJ000405.10-165951.5 & 0.60607392 & 0.6$\pm$0.2 & 12.4$\pm$0.1 & UU Cet  &  \\ 1SWASPJ000409.77-410810.8 & 0.52548796 & 1.0$\pm$0.6 & 14.5$\pm$0.5 &  & J000409.8-410811 \\ 1SWASPJ000543.68-343400.7 & 0.47432521 & 0.3$\pm$0.6 & 15.1$\pm$0.4 &  & J000545.1-343445 \\ 1SWASPJ000548.23+424321.9 & 0.55597466 & 0.6$\pm$0.4 & 13.8$\pm$0.2 &  &  \\ 1SWASPJ000552.59-335119.2 & 0.53367919 & 0.8$\pm$0.9 & 15.7$\pm$0.7 &  & J000552.6-335120 \\
\hline

\end{tabular}
\tablefoot{$A_{\rm{LC}}$ refers to the peak to peak amplitude of the pulsation in magnitudes and $\bar{m}_{\rm{LC}}$ denotes the median magnitude of the light curve. The SSS / CSS name refers to the unique identifier from the Siding Springs Survey or the Catalina Schmidt Survey if the SSS name is not known. The full catalogue of 4963 objects is available at CDS.}
\end{table*}

\section{The Blazhko effect in the frequency domain }
\label{sec:FreqDomain}
\subsection{Frequency domain investigation}
In addition to the main pulsation signal and its higher frequency harmonics, we anticipated that the Blazhko effect could be detected as modulation sidebands, and/or as a low frequency peak in the power spectra of SWASP light curves.
Low frequency peaks at the Blazhko frequency itself are not often detected in conventional observations of RR Lyrae type stars, owing to the limited baselines usually available. The unique feature of SWASP data are their long baselines and high cadence, meaning that there is chance to detect Blazhko modulations directly in the power spectrum. Sidebands, which are pairs of peaks on either side of the main pulsation peak, can be created by both frequency (FM) and amplitude (AM) modulation \citep{bracewell_fourier_1965}, and modulation has been observed forming triplets and quintuplets of peaks in high--resolution Blazhko effect studies \citep{benko_blazhko_2011}. The difference between the main pulsation peak and each sideband provides the modulation frequency, in this case the Blazhko effect frequency  $\left( f_{\rm{BL}} \right)$: $f_{\rm{BL}} = f_{0} - f_{-}$ , or $f_{\rm{BL}} = f_{+} - f_{0} $
where $f_{0}$ is the main pulsation frequency and $f_{-}$ and $f_{+}$ are the frequencies of the lower and upper sidebands respectively. However, some care is still needed as sidebands can also be caused by systematic periodic variation in flux due to diurnal and synodic cycles in the case of ground--based observations. The combination of AM, FM and irregular sampling causes asymmetry in sideband amplitude, sometimes to the extent that one component may be missing entirely.

Therefore, in order to counteract the spectral effects due to ground--based observations, an investigation of the power spectra of the SuperWASP RRab objects was carried out using a variable gain implementation of the {\sc CLEAN} algorithm \citep{roberts_time_1987,hogbom_aperture_1974} which deconvolves the signal from the complex window function created by irregularly sampled ground based observations. It is particularly well suited to light curves containing multiply period signals. {\sc CLEAN}ed power spectra were therefore calculated from the light curves of each of the 4963 RRab objects, with flux values in SWASP counts, where a count rate of 1 c s$^{-1}$ corresponds to $m_V\sim15$. The parameters of {\sc CLEAN} were set to allow the maximum resolution of the power spectra to be calculated based on the duration of observations. The {\sc CLEAN} algorithm first calculates the window function of the light curve based on the sampling regime. It then convolves this window function with a delta function corresponding to a certain percentage (defined by the gain) of the highest peak in the power spectrum. The result of this convolution is then subtracted from the actual power spectrum. This process is then repeated until a pre-defined maximum number of iterations has been reached or there is no further change to the power spectrum. The accumulated delta functions used are then recombined with the residual power spectrum from the last iteration to produce the {\sc CLEAN}ed spectrum. In this way the pre--whitening of the spectrum is performed gradually on all peaks in the power spectrum while removing the majority of the effects of aliasing. Peaks maintain a large proportion of their power while the rest of the spectrum is reduced to the underlying noise level.

In order to identify the specific noise level in the power spectrum for each object after {\sc CLEAN}ing, a spectral representation of a non--varying version of that object was obtained using the object's noise, average brightness and spectral window. Removal of the variability was achieved by creating a copy of each object's light curve with the actual fluxes randomly reassigned to the times of observation. These randomised light curves were also {\sc CLEAN}ed with the same parameters as the actual observations in order to duplicate the removal of the average brightness and number of {\sc CLEAN} iterations that the actual signal underwent. Since there is no signal to convolve with the window function and then be {\sc CLEAN}ed out, the power spectrum of the randomised light curve (labelled {\sc CLEAN}ed RDM in fig~\ref{fig:LFPspec} and \ref{fig:Dayspec}) is therefore a good indicator of the noise level in the power spectrum of a non--varying object, against which to measure the signal in all regions of the actual {\sc CLEAN}ed power spectrum.

To identify a candidate Blazhko period from a {\sc CLEAN}ed power spectrum, we required at least one of the following matching criteria to be satisfied. Either there was a significant low frequency peak in the power spectrum whose frequency matched that determined from a significant (positive or negative) sideband to the main pulsation frequency, and/or there was a symmetrical pair of significant positive and negative sideband peaks to the main pulsation frequency peak. Identification of candidate Blazhko periods therefore focused on two frequency ranges and required us to define a noise level in the power spectrum for each region. For low frequency peaks in the region of the Blazhko effect signal itself, the noise level was calculated between $3.17\times10^{-8}$~Hz and $2.3{\times}10^{-6}$~Hz (corresponding to periods between 1 year and 5 days). The limit of 1 year was made to ensure that observations covered several cycles during most SWASP observations and to reduce the likelihood of an annual systematic signal being mistaken for a Blazhko period. The noise level in the vicinity of the sidebands was calculated between $7.7{\times}10^{-6}$~Hz and $10^{-4}$~Hz, equating to periods between 1.5~d and 0.1~d. In both the low frequency and sideband frequency range, an initial first order polynomial was fitted to the power spectrum of the randomised data to define an initial noise baseline. Any peaks in the power spectrum of the randomised data lying above 4 times this first baseline were removed, and the remaining power spectrum of the randomised data was then fitted again with a first order polynomial to determine a final noise level. 
Thresholds for acceptance of a peak in the power spectrum of the actual data were set at 4 times this noise level for low frequency peaks and at 10 times this noise level for sidebands. The higher sideband limit was set in order to avoid the higher level of noise and spurious peaks surrounding the base of the main frequency without excluding any genuine sidebands. As described above, rather than finding many harmonic orders of side peaks around the main pulsation frequency, the aim was to detect a strong sideband to link with a low frequency peak, or a pair of symmetrical sidebands.
Only those peaks in the power spectrum of the actual data which exceeded these multiples of the noise level were considered genuine. Peaks which did exceed these levels were then each fitted with a Gaussian curve in order to yield a measurement of the peaks' maximum power, central frequency and the uncertainty in frequency, based on the Gaussian width.

As already noted, candidate Blazhko periods were defined for those objects whose sideband peak's separation from the fundamental mode peak matched a low frequency peak, and/or where symmetrical pairs of modulation sidebands existed on either side of the fundamental mode. In both cases a match was defined to exist where the 1-sigma limits of each measured frequency overlapped with each other. In other words, at least two peaks at the same modulation frequency within a $1\sigma$ uncertainty were required for identification of Blazhko candidates.

Following their identification from the {\sc CLEAN}ed power spectra, all candidate Blazhko frequencies were then refined using the Period04 program \citep{lenz_period04_2005} by fitting sinusoids (individually in the case of several Blazhko periods per object) to their respective object's light curve. The resulting Blazhko periods and amplitudes, along with the analytical uncertainties from Period04, are given in Table~\ref{tab:blazcat}.

\subsection{{\sc CLEAN} results}
\label{CLEANresults}
\begin{figure}[h]
 \centering
   \resizebox{\hsize}{!}{\includegraphics{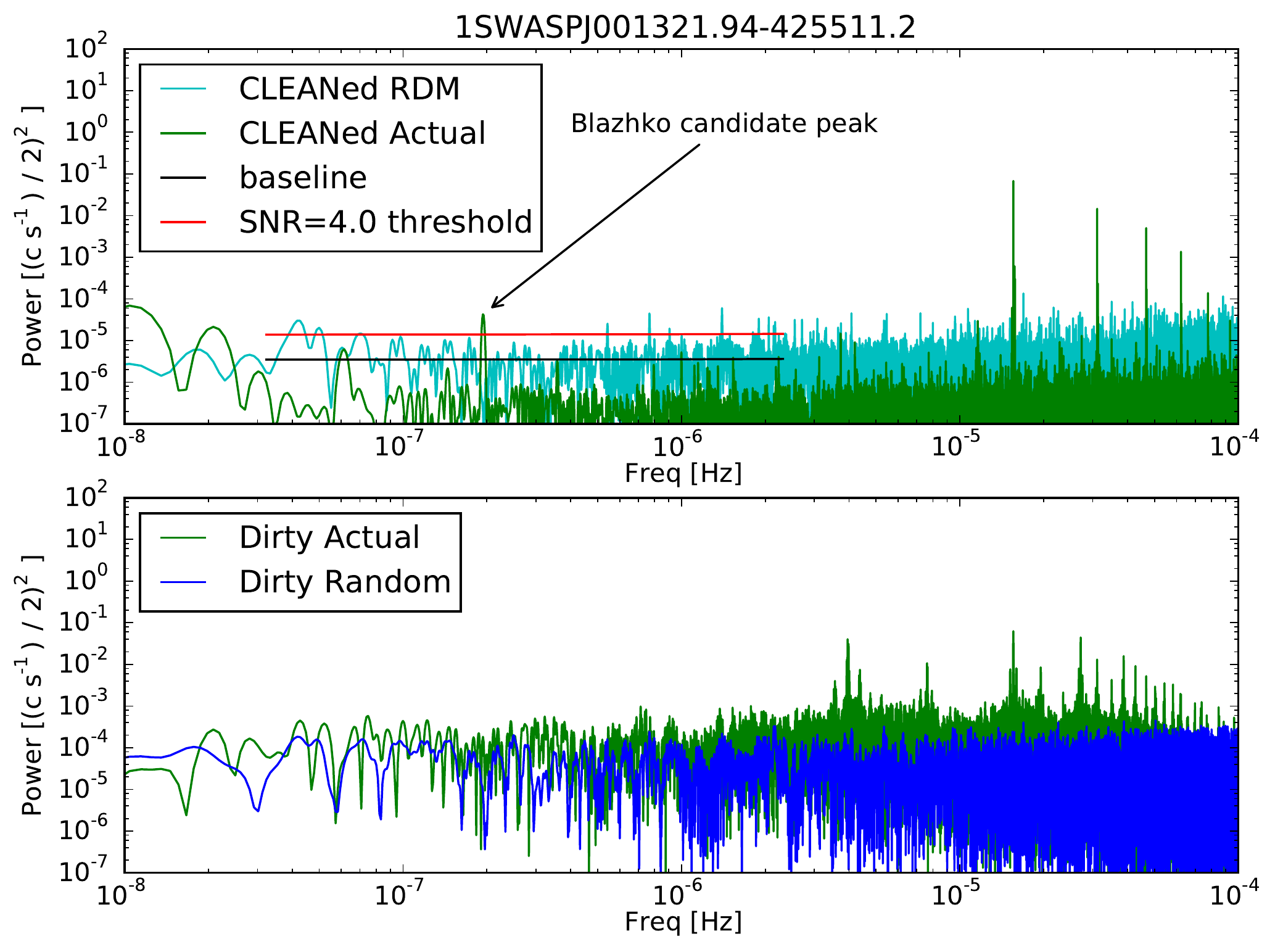}}
      \caption{{\sc CLEAN}ed power spectrum of \object{1SWASPJ001321.94-425511.2} from $10^{-8}$~Hz to $10^{-4}$~Hz. The top panel shows the power spectra of the original and randomised light curves after deconvolution with the {\sc CLEAN} algorithm whilst the bottom panel shows the raw (or "dirty") power spectra of the actual and randomised data. Peaks due to the actual signal can be seen more clearly once noise due to the convolution of the signal with the window function has been reduced. The red line in the top panel is the noise threshold calculated for frequencies below $2.3{\times}10^{-6}$~Hz (equating to Blazhko periods of at least 5 days).}
   \label{fig:LFPspec}
   \end{figure}

\begin{figure}[h]
  \centering
    \resizebox{\hsize}{!}{\includegraphics{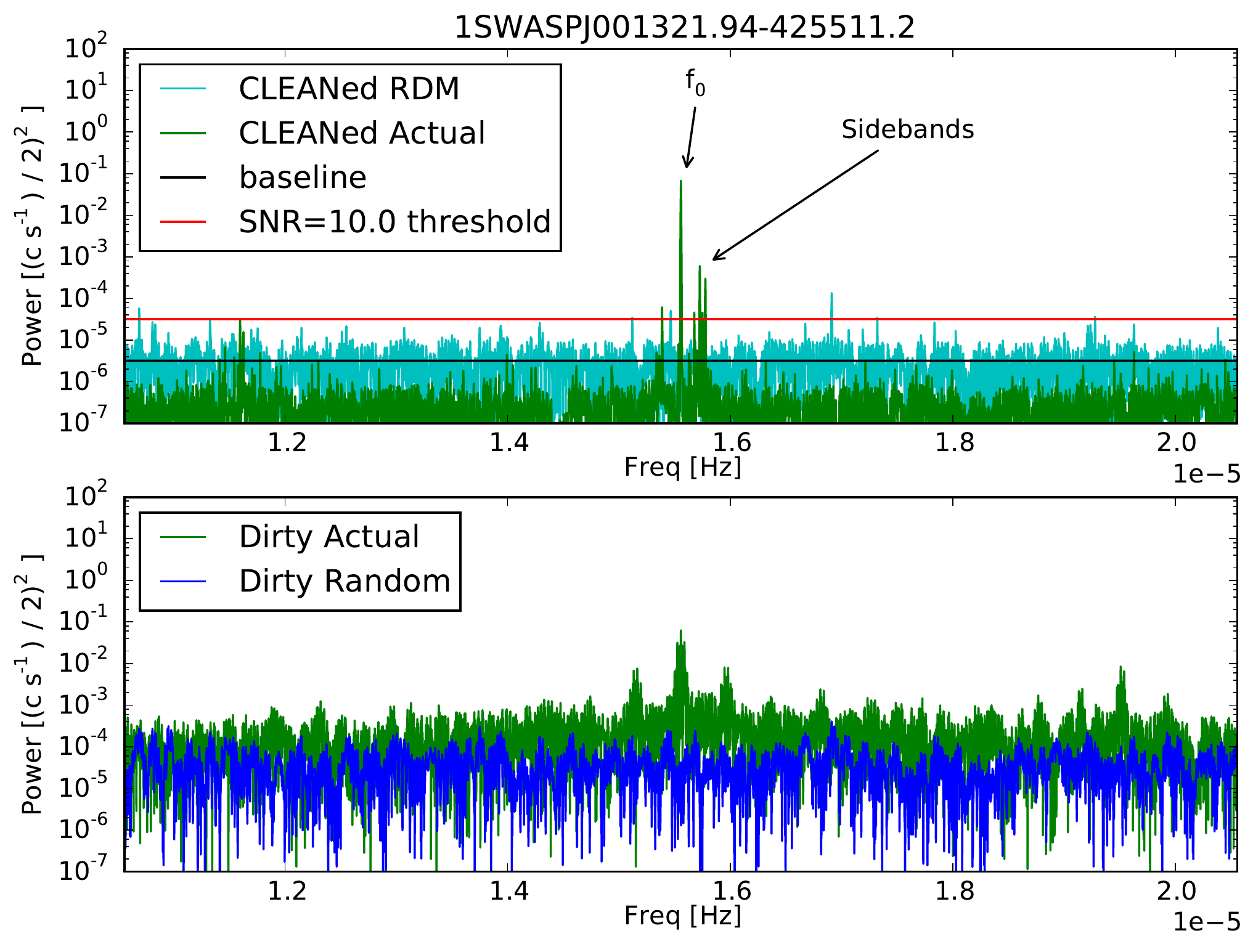}}
      \caption{{\sc CLEAN}ed power spectrum of \object{1SWASPJ001321.94-425511.2} focusing on the range $5\times10^{-6}$~Hz on either side of the main pulsation peak at $1.56\times10^{-6}$~Hz (1.344 c d$^{-1}$). The panels are arranged as in Fig.~\ref{fig:LFPspec}. The main pulsation peak, $f_0$, and some of the modulation sidebands on the right side of the main pulse frequency have been labelled.}
   \label{fig:Dayspec}
   \end{figure}

As shown in Figs.~\ref{fig:LFPspec} and \ref{fig:Dayspec}, the {\sc CLEAN} algorithm proved to be excellent at deconvolving the window function from the actual signal allowing low frequency periodic signals to appear above the baseline noise level defined by the fit to the randomised noise signal. This was particularly useful when distinguishing Blazhko candidate peaks in the low frequency range described above. The "dirty" power spectrum of the light curve and corresponding randomised spectra are shown in the bottom panel of Figs.~\ref{fig:LFPspec} and \ref{fig:Dayspec} for comparison. 

\begin{figure*}[h]
  \centering
   \subfloat[][1SWASPJ000548.23+424321.9]{\includegraphics[width=7cm]{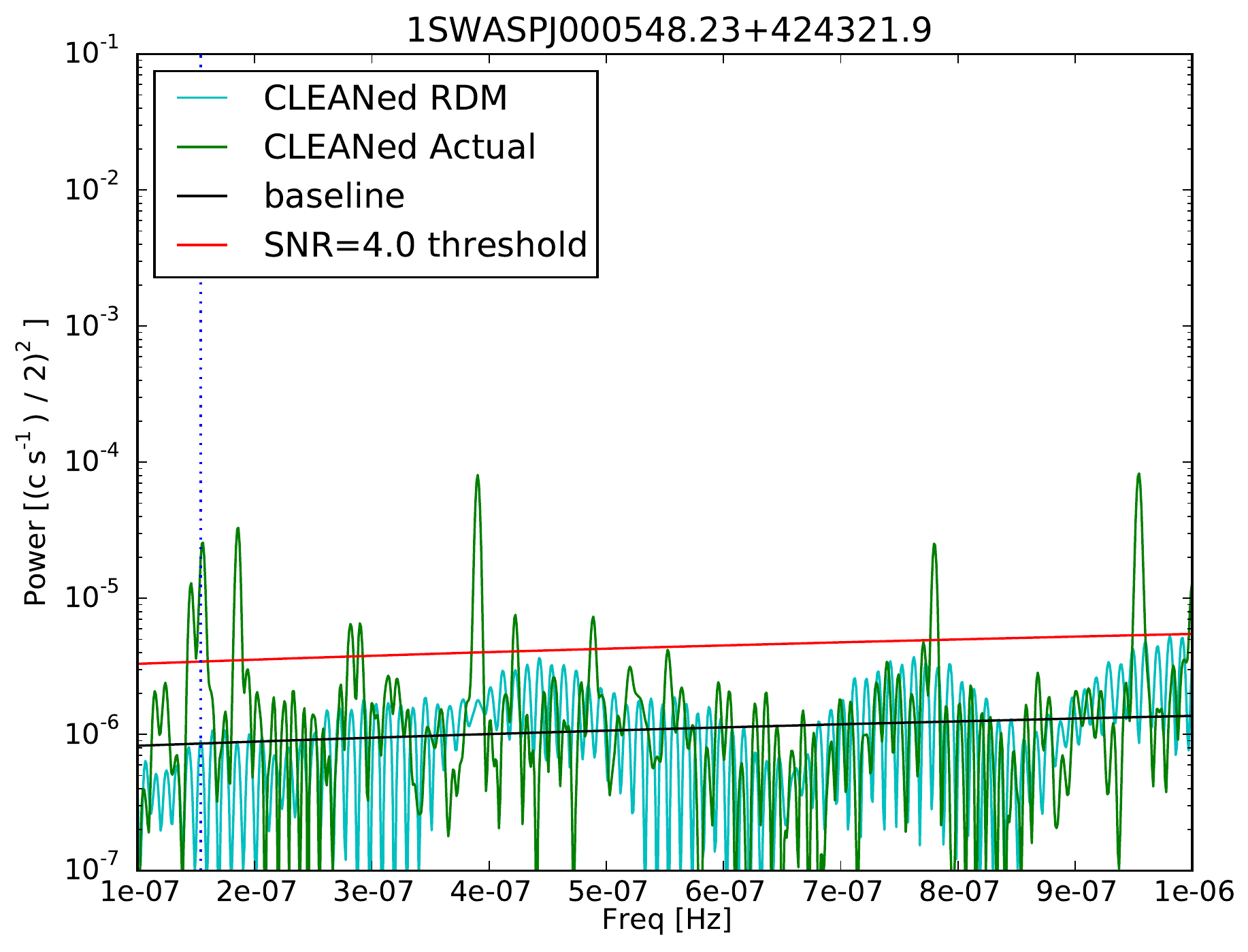}\label{fig:548LFPspec}}
    \subfloat[][1SWASPJ000548.23+424321.9]{\includegraphics[width=7cm]{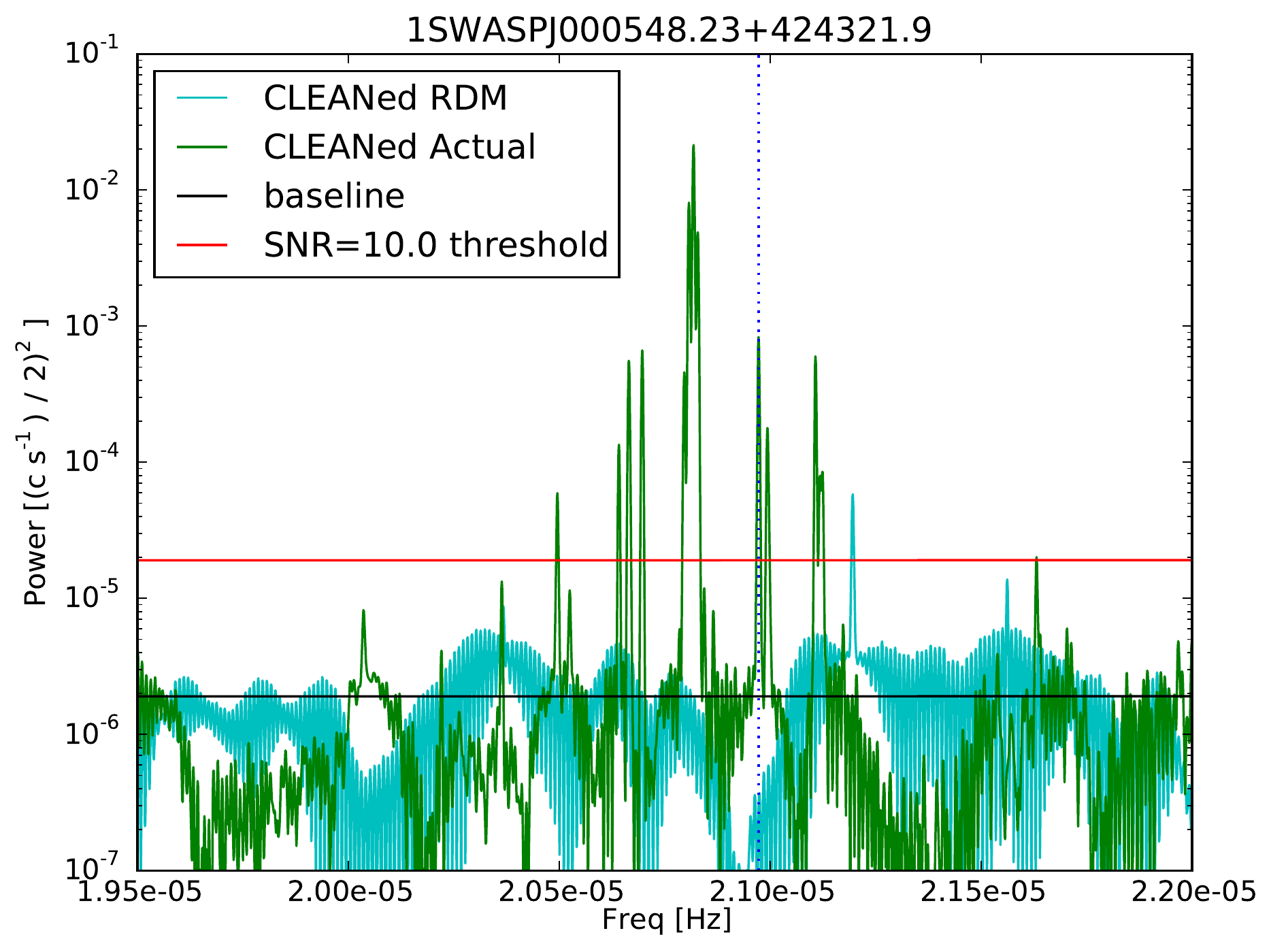}\label{fig:548Dayspec}}
      \caption{Sections of the {\sc CLEAN}ed power spectra showing in greater clarity the matching low frequency peaks and sidebands for \object{1SWASPJ000548.23+424321.9}, highlighted by the vertical dotted lines. As before, the randomised signal is pale blue with the actual signal in green. The baseline created using this randomised signal is in black, with the signal to noise threshold at 4 times this shown in red. The resultant $\rm{P_{BL}}$ is 74.8 d.}
   \label{fig:548zoom}
   \end{figure*}

\begin{figure*}[h]
 \centering
   \subfloat[][1SWASPJ001321.94-425511.2]{\includegraphics[width=7cm]{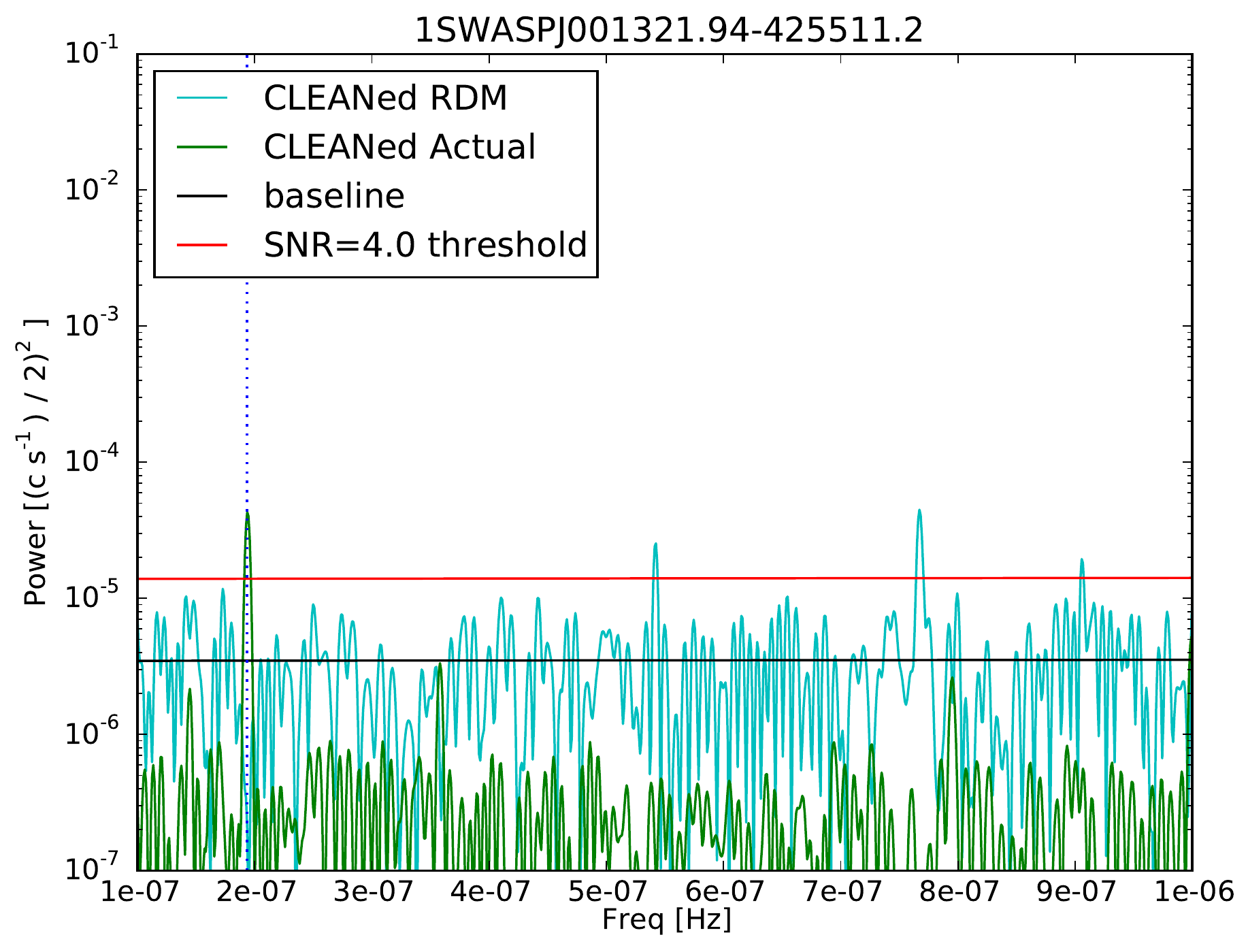}\label{fig:1321LFPspec}}
    \subfloat[][1SWASPJ001321.94-425511.2]{\includegraphics[width=7cm]{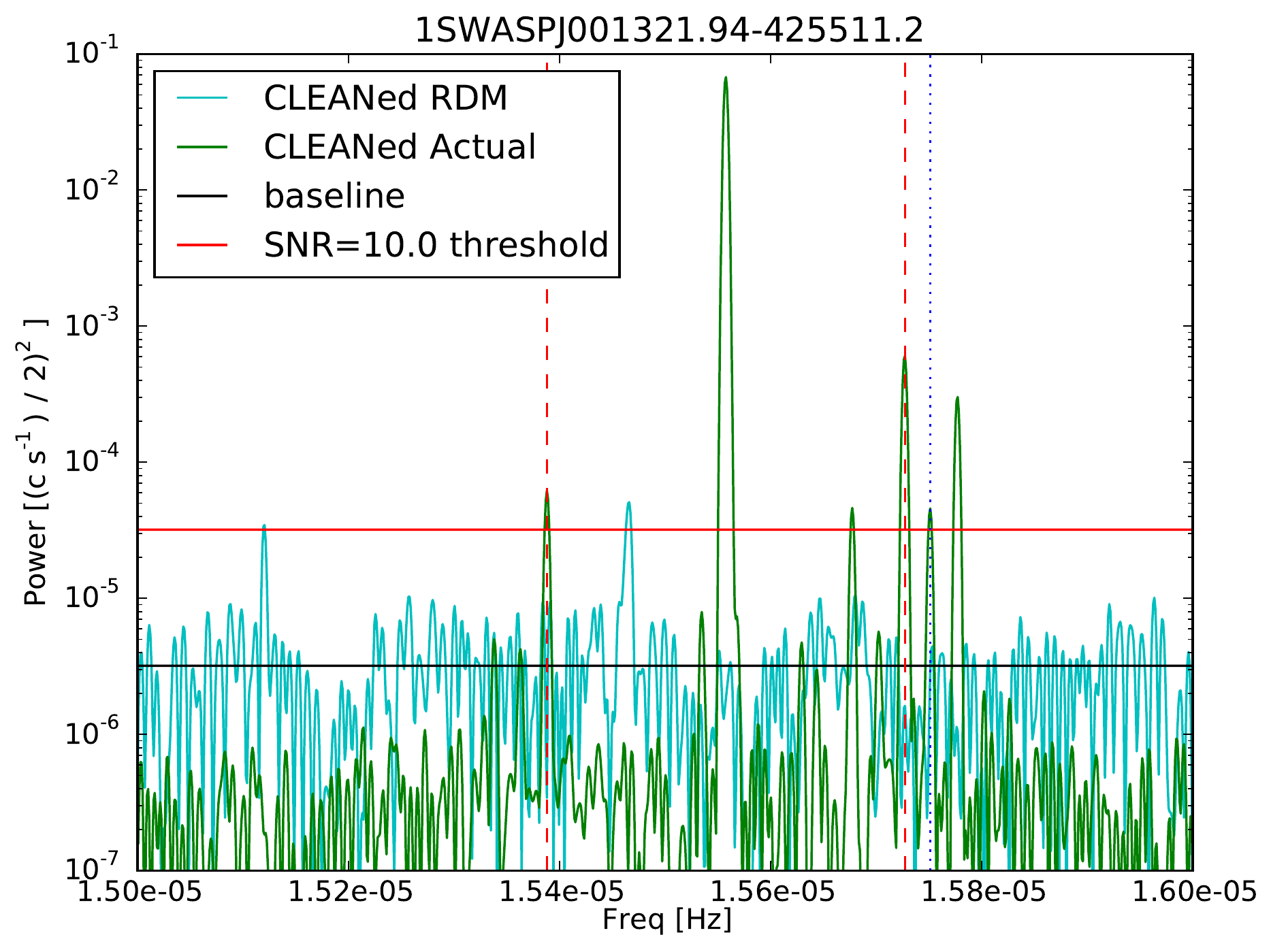}\label{fig:1321Dayspec}}

      \caption{Sections of the {\sc CLEAN}ed power spectra showing in greater clarity the matching low frequency peaks and sidebands for \object{1SWASPJ001321.94-425511.2}. The colour scheme is the same as in Fig.~\ref{fig:548zoom}. The pair of matching sidebands ($\rm{P_{BL}}$=68.3 d) are marked with dashed vertical lines, whereas the low frequency peak--sideband pair ($\rm{P_{BL}}$=59.8 d) are marked with dotted vertical lines.}
   \label{fig:1321zoom}
   \end{figure*}

Close--ups of the low frequency peaks and areas of the spectra around the main pulsation frequency containing the modulation sidebands are shown in Fig.~\ref{fig:548zoom} for \object{1SWASPJ000548.23+424321.9} and Fig.~\ref{fig:1321zoom} for \object{1SWASPJ001321.94-425511.2}. The matching sidebands and low frequency peaks are highlighted by the blue dotted line for clarity. In the case of \object{1SWASPJ001321.94-425511.2}, one Blazhko period was discovered based on a pair of equidistant sidebands, marked with dashed lines, and the second Blazhko period was discovered by a matching low frequency peak and sideband, highlighted with dotted lines again.

As a result of the analysis described above, a disproportionate number of Blazhko candidate periods were detected below 10~d. This appears to be due to a dominance of systematic aliases at around 5~d and 7~d, most likely due to the sampling regime. Since only 4\% of Blazhko periods are below 10~d in the online Skarka database, the decision was taken to exclude Blazhko candidates below 10~d in our Blazhko catalogue. Furthermore, 356 objects had an excessive number of candidate low frequency peaks and they too were excluded from further consideration.

\begin{figure}[ht]
  \begin{center}
   \resizebox{\hsize}{!}{\includegraphics{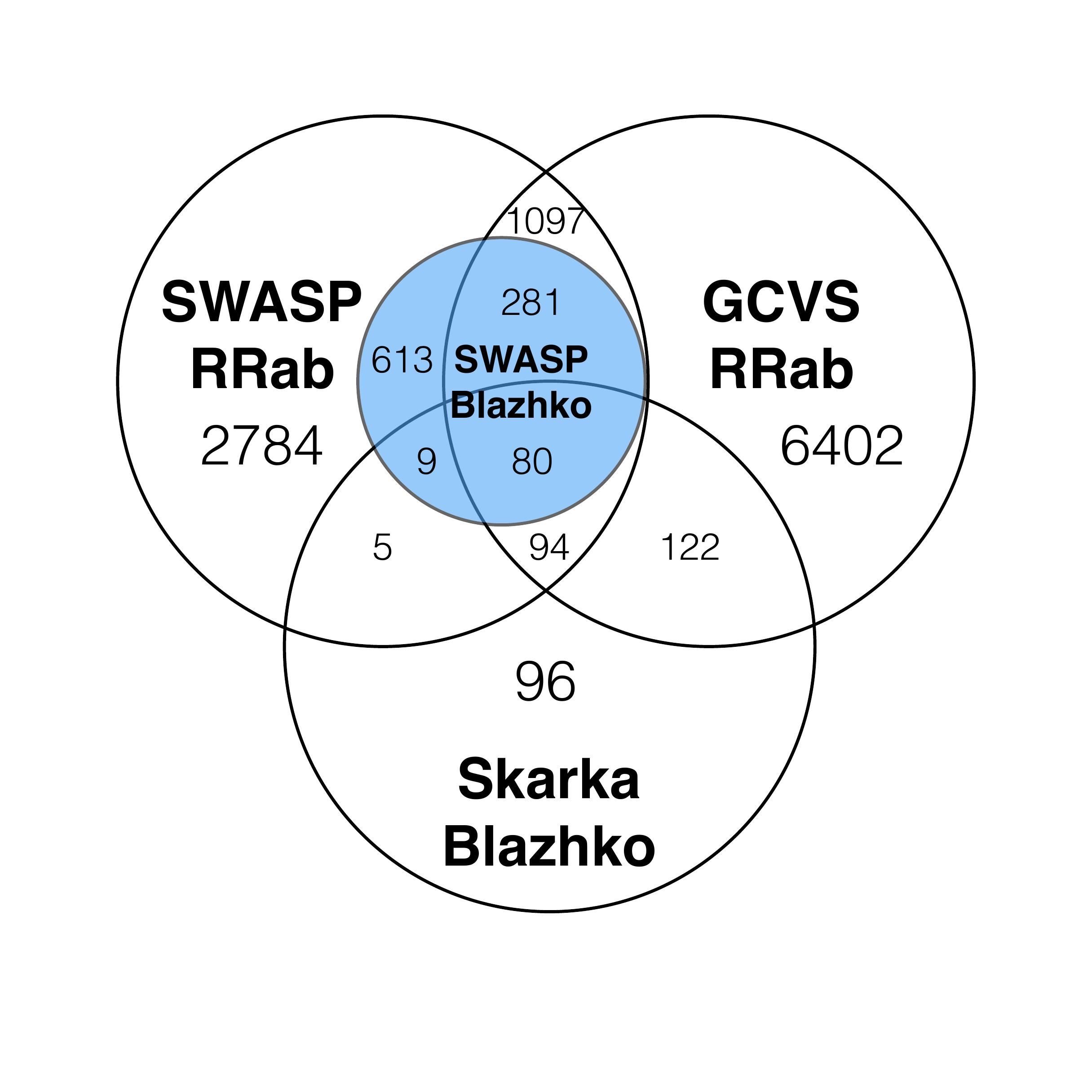}}
  \caption{Quantities of objects from the 3 catalogues used in this study. The 983 Blazhko objects discovered using the {\sc CLEAN} routine are labelled "SWASP Blazhko". The 613 newly identified Blazhko objects are distinguished from those found in existing lists.}
  \label{fig:Venn}
  \end{center}
\end{figure}

Following these cuts, a total of 983 candidate Blazhko effect stars were identified from the SuperWASP sample of RR Lyrae. This is 19.8\% of the 4963 RRab objects in the SWASP RRab catalogue. Of these candidate Blazhko objects, 89 were previously known Blazhko effect stars listed by Skarka. This leaves 894 Blazhko stars newly discovered in the SuperWASP archive. 
The 983 Blazhko candidates had a total of 1386 Blazhko periods, with 272 objects having more than one potential Blazhko period. 
Multiplets, such as triplets or quintuplets, of sidepeaks can produce sets of modulation frequencies at integer multiples of the actual modulation frequency. Each candidate Blazhko object with multiple Blazhko periods was checked to see if any periods were small integer multiples of each other, and then if any remaining periods were duplicates (due to multiple CLEAN periods resulting in the same Period04 period).
31 harmonic periods were found, including \object{1SWASPJ132305.17-205939.0} which had periods in the ratio of 2~:~3~:~6.  
These small integer multiples were removed, along with any duplicate periods resulting a total number of independent Blazhko periods of 1324 for the same 983 objects. Our results still show 261 objects with more than 1 independent Blazhko period.
1552 of the SWASP RRab catalogue objects match objects of the same type in the GCVS catalogue, amongst which are 361 of our Blazhko candidates. These figures are represented in Fig.~\ref{fig:Venn}.

Existing studies have focused on identifying the Blazhko effect using modulation sidebands only, whereas we also allowed Blazhko candidates identified from the presence of a low frequency peak. However, we emphasize that we required a match between either a low frequency peak and a sideband, or a pair of sidebands, before we accepted an object as a candidate. So two matching signals in the power spectrum were required for each candidate object. There are 99 known Blazhko effect stars that are in the list of SuperWASP RRab objects which were not identified as Blazhko using the processes described above, possibly because of the strict criteria we adopted.

The first 20 rows of the Blazhko catalogue sorted by right ascension are listed in Table~\ref{tab:blazcat}, showing the pulsation amplitude $A_{\rm{LC}}$ and the amplitude of the Blazhko effect peak as $A_{\rm{BL}}$, both in magnitudes. Where the RRab object matched a Skarka object, their Skarka name is listed along with any previously known Blazhko periods.

For completeness, the catalogue of Blazhko candidates was compared, using the CDS Xmatch service, to the \cite{drake_catalina_2014} Periodic Variables catalogue of 47,055 objects, which contains 73 objects flagged as Blazhko. This cross-match found 35 SWASP candidate Blazhko objects in the \cite{drake_catalina_2014} Blazhko objects, 3 of which appear in the Skarka on-line database. There are 157 Blazhko objects listed in the Drake catalogue that were not detected as Blazhko in our SWASP RRab catalogue, 6 of which appear in the Skarka list. 

\begin{sidewaystable*}[ht]
	\caption{Blazhko candidate results}
	\label{tab:blazcat}
\centering
\begin{tabular}{l l l l l l l } \hline \hline
 SWASPid & $P_{\rm{Pulse}}\ $ [s] & $A_{\rm{LC}}$ [mag] & $P_{\rm{BL}}$ [d] & $A_{\rm{BL}}$ [mmag] & Skarka name & Skarka $P_{\rm{BL}}$ [d]   \T  \\
 \hline
1SWASPJ000035.58+263949.6 & 0.56606638 & 0.750 & 120.0$\pm$0.3 & 106 & GV Peg & 80 \\ 1SWASPJ000049.23-284900.4 & 0.58303469 & 0.479 & 73.3$\pm$0.1 &  34 &  &  \\ 1SWASPJ000126.73-001344.2 & 0.55574918 & 0.544 & 155.2$\pm$0.8 &  23 &  &  \\ 1SWASPJ000157.75-364042.4 & 0.63752764 & 0.454 & 27.0$\pm$0.0;26.6$\pm$0.0 &  37; 52 & J000157.75-364042.4 & ... \\ 1SWASPJ000409.77-410810.8 & 0.52548796 & 0.968 & 201.6$\pm$0.8 &  46 &  &  \\ 1SWASPJ000548.23+424321.9 & 0.55597466 & 0.557 & 74.8$\pm$0.1 &  59 &  &  \\ 1SWASPJ000552.59-335119.2 & 0.53367919 & 0.788 & 42.6$\pm$0.1;46.0$\pm$0.1 &  10; 14 &  &  \\ 1SWASPJ000624.38+170443.3 & 0.53171468 & 0.755 & 340.5$\pm$3.4 &  33 &  &  \\ 1SWASPJ000652.94+114114.0 & 0.58486086 & 1.002 & 52.1$\pm$0.2;285.8$\pm$1.5 &  19; 67 &  &  \\ 1SWASPJ000823.86+263323.3 & 0.53028244 & 0.675 & 37.1$\pm$0.0;28.4$\pm$0.0;282.6$\pm$1.4 &  35; 22; 58 &  &  \\ 1SWASPJ001033.40-142439.9 & 0.60850245 & 1.029 & 125.9$\pm$0.7 &  44 &  &  \\ 1SWASPJ001321.94-425511.2 & 0.74395519 & 0.548 & 59.4$\pm$0.1;67.6$\pm$0.1 &  29; 20 &  &  \\ 1SWASPJ001333.20+282942.9 & 0.47242147 & 0.409 & 53.6$\pm$0.1 &  22 &  &  \\ 1SWASPJ001334.56-095019.3 & 0.56786352 & 0.572 & 33.6$\pm$0.2 &   4 &  &  \\ 1SWASPJ001408.92-092905.4 & 0.61616677 & 0.893 & 27.2$\pm$0.0 &  31 &  &  \\ 1SWASPJ001512.51-351541.6 & 0.60572129 & 0.694 & 78.5$\pm$0.4 &   6 &  &  \\ 1SWASPJ001748.57+095322.3 & 0.70111263 & 0.876 & 314.8$\pm$3.3 & 242 &  &  \\ 1SWASPJ002322.79+134541.0 & 0.53127819 & 1.033 & 163.6$\pm$0.4;290.7$\pm$1.1 & 359;469 & FI Psc & >120 \\ 1SWASPJ002521.68-363024.1 & 0.69991457 & 0.744 & 36.0$\pm$0.1 &   9 &  &  \\ 1SWASPJ002843.08-440022.6 & 0.59996766 & 0.765 & 117.9$\pm$0.2;117.9$\pm$0.2 & 135;135 &  &  \\
\hline
\end{tabular}
\tablefoot{The full catalogue of 983 objects is available at CDS.}
\end{sidewaystable*}

\section{The Blazhko Effect in the Time Domain}
\label{sec:TimeDomain}
\subsection{Quantifying the Blazhko effect using PDM}
\begin{figure}[ht]
\centering
  \resizebox{\hsize}{!}{\includegraphics{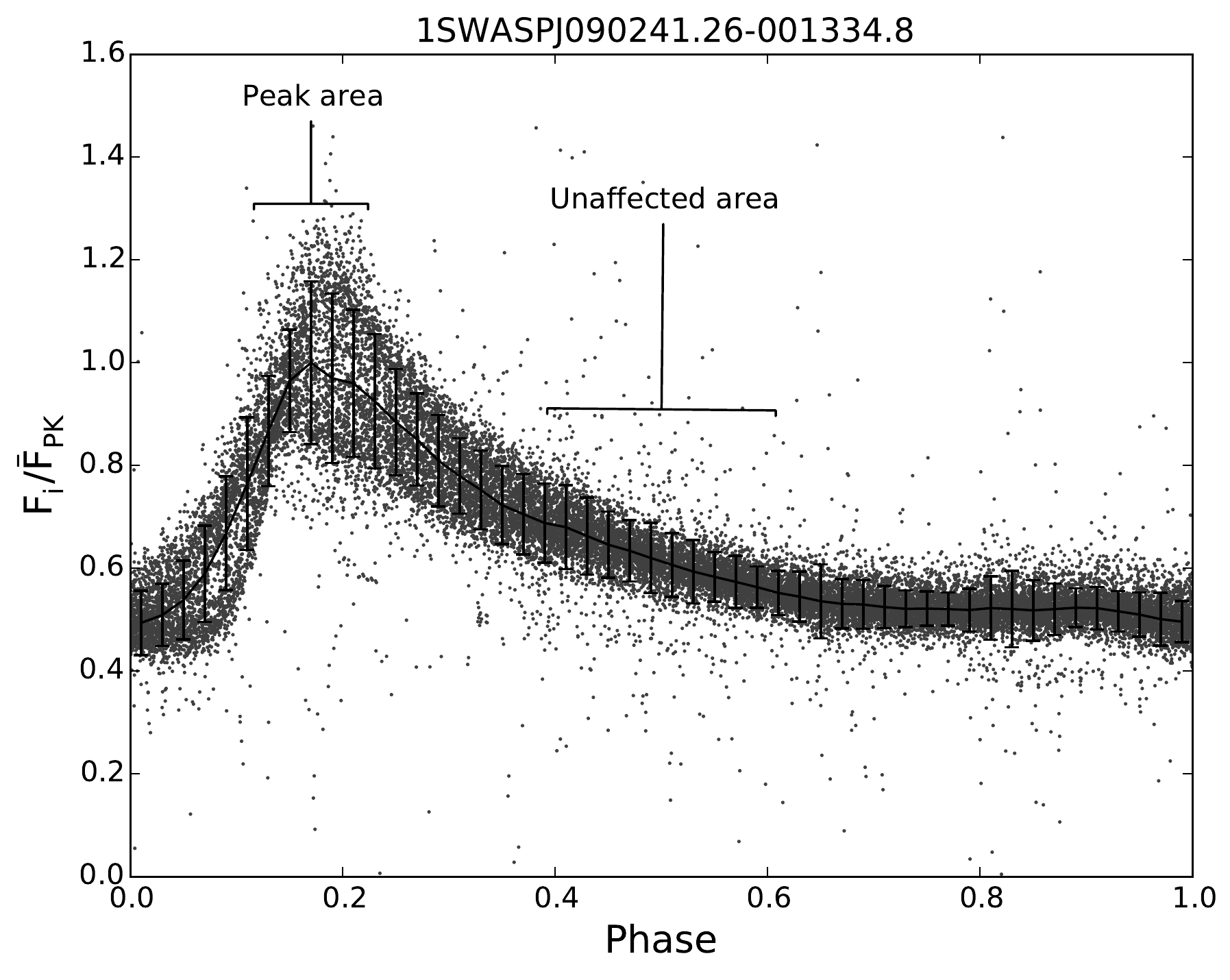}}
  \caption{Phase folded light curve for \object{1SWASPJ09024.26-001334.8} showing the areas used to create the relative scatter parameter, labelled "Peak area" and  "Unaffected area". The median flux level for each bin is shown in blue with 1$\sigma$ error bars. The vertical axis is in flux units relative to the peak bin median flux level.}
  \label{fig:AnnotatedLC}
\end{figure}

In initial phase folding tests using 12 objects from Skarka's list of known Blazhko effect stars \citep{skarka_blasgalf_2013} the scatter at the peak of the pulsation was seen to be more pronounced than elsewhere in the light curves, which seemed to be a symptom of AM, as noted by \cite{skarka_bright_2014}. \cite{payne_identification_2013} also noticed this anomaly and suggested considering the level of scatter of the light curve as a "Blazhko Indicator". Hence, in order to compare levels of scatter between different light curves, the phase folded light curves were normalised by dividing each curve's flux counts by its peak bin median flux and phase shifted so the minimum corresponds to the first bin. We now define the envelope function as the variation in the peak and minimum of the light curve. We calculated the ratio of noise in the peak area to the noise in a quiet part of the descending section of the light curve in order to give a relative scatter parameter. Using the same 50 bins as the PDM routine, this quiet area is from bins 20 to 29 and labelled as "Unaffected area" in Fig.~\ref{fig:AnnotatedLC} because the amount of noise in this descending section appeared to be dependent only on the overall quality of the light curve, and not on AM.
With the light curve split into 50 bins the highest 3 bins of the folded light curves were defined as the peak area as shown in Fig.~\ref{fig:AnnotatedLC}.

The standard deviation in each of the bins, $\sigma_{bin}$, is defined as
\begin{equation}
 \sigma_{bin} = \left(\frac{ \Sigma_{i=1}^{i=N_{bin}}{\left(F_{i} - \tilde{F}_{\rm{bin}}\right)^2} }{ \left(N_{bin}-1\right)} \right)^{1/2} \nonumber
	\label{eq:sd}
\end{equation}
where $F_{i}$ is the $i$-th flux point in the bin, $\tilde{F}_{\rm{bin}}$ is the median flux in the bin, and $N_{bin}$ is the total number of flux points in the bin.
The relative scatter was quantified as  $\sigma_r=\sigma_p / \sigma_q$ where $\sigma_p$ is the mean of $\sigma_{\rm{bin}}$ across the 3 peak bins and $\sigma_q$ is the mean of $\sigma_{\rm{bin}}$ across bins 20 to 29.

\begin{figure}[ht]  
	\def\tabularxcolumn#1{m{#1}}
	\begin{tabularx}{\hsize}{@{}cXX@{}}
	\begin{tabular}{ccc}
\subfloat{\includegraphics[width=0.32\hsize]{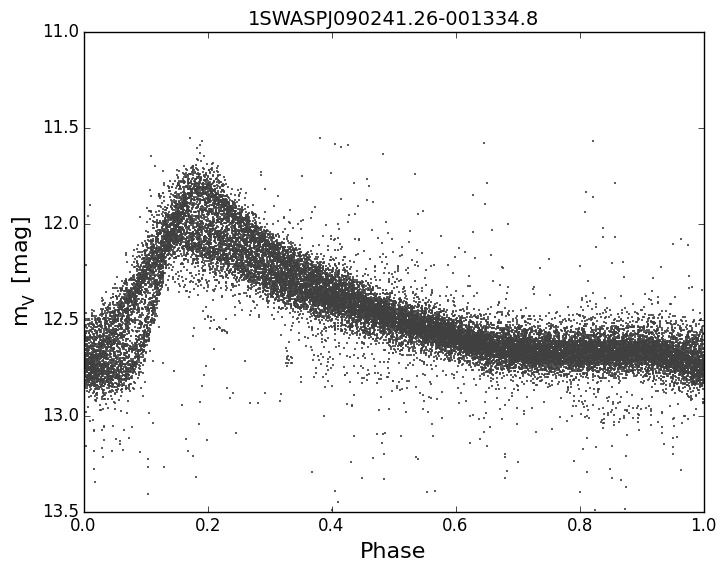} }
    \subfloat{\includegraphics[width=0.32\hsize]{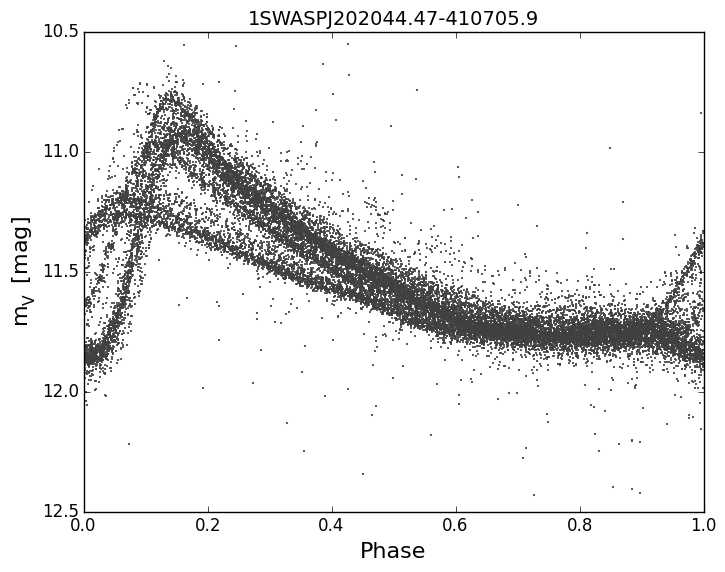}}
    \subfloat{\includegraphics[width=0.32\hsize]{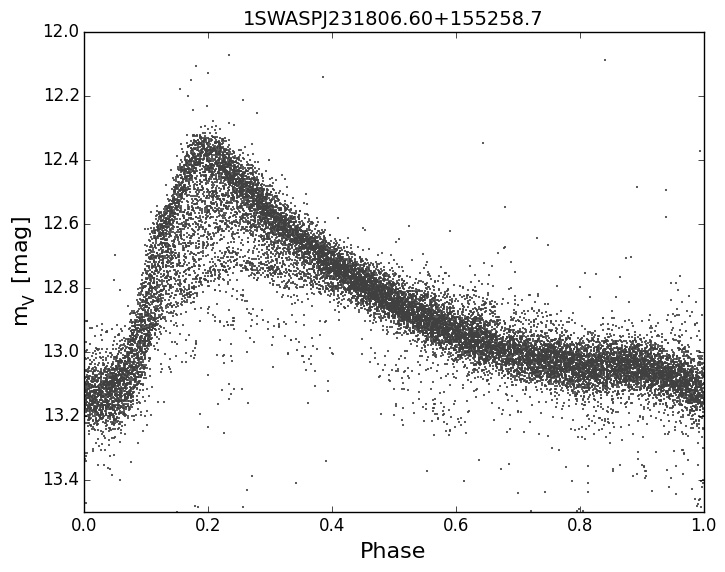}}\\
\subfloat{\includegraphics[width=0.32\hsize]{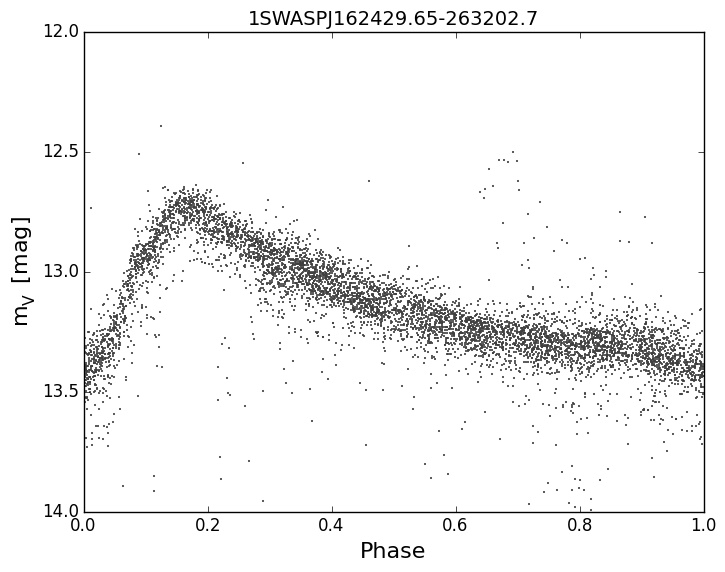}}
   \subfloat{\includegraphics[width=0.32\hsize]{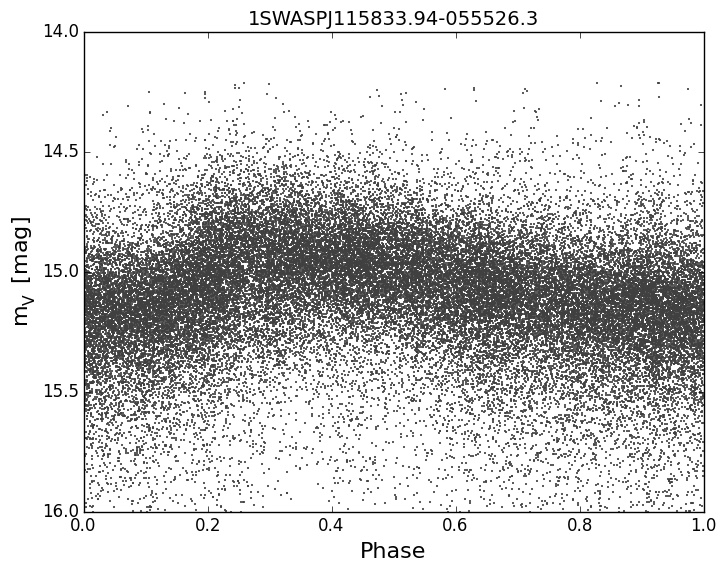} }
   \subfloat{\includegraphics[width=0.32\hsize]{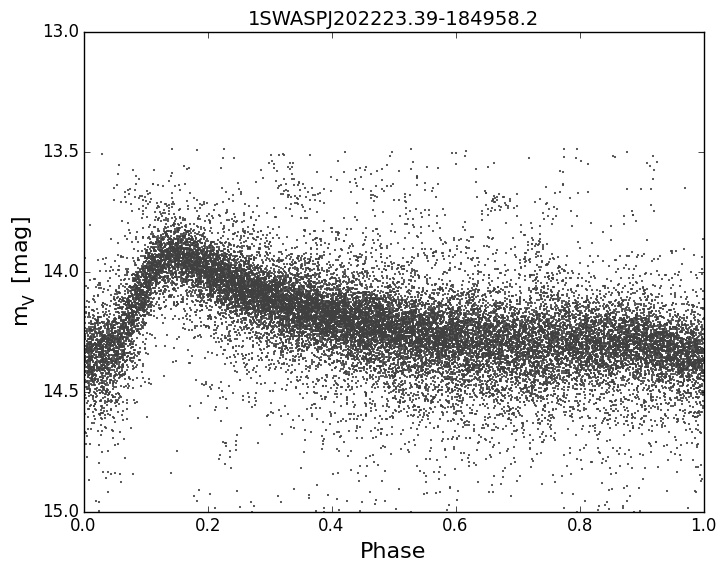}}\\
\subfloat{\includegraphics[width=0.32\hsize]{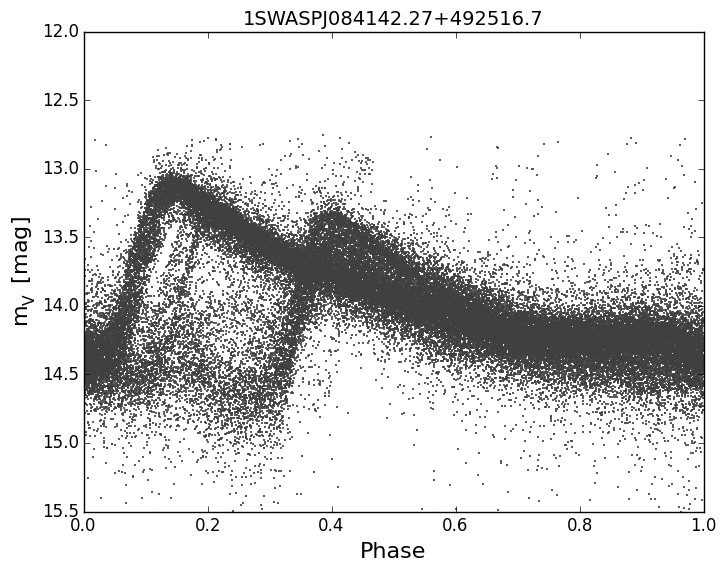}}
  \subfloat{\includegraphics[width=0.32\hsize]{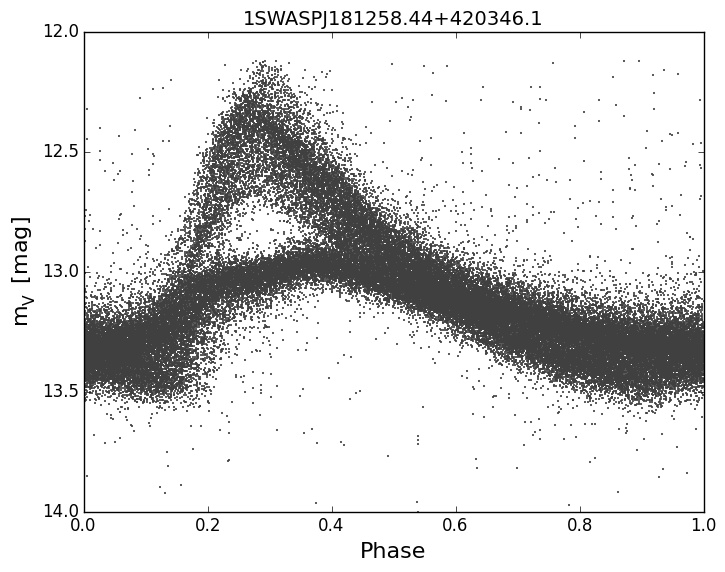}}
  \subfloat{\includegraphics[width=0.32\hsize]{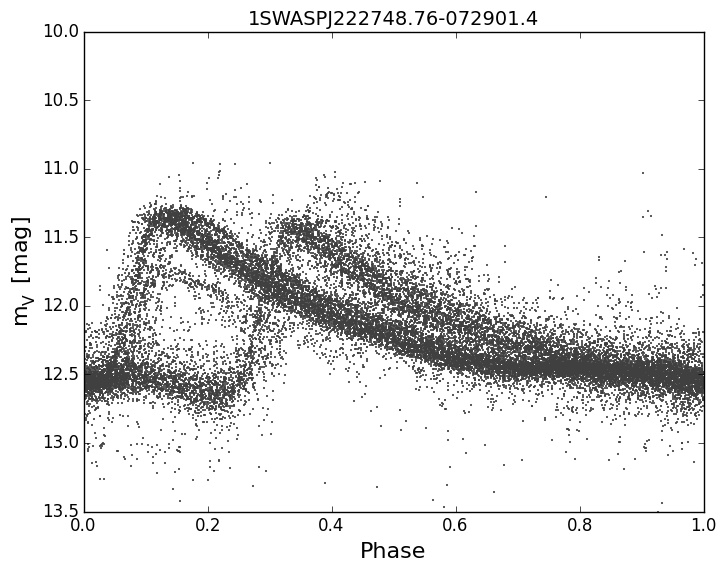}}
	\end{tabular}
	\end{tabularx}
\caption{Phase folded light curves for Blazhko Catalogue objects. The vertical scale is in magnitudes. The top row shows objects identified as Blazhko effect objects and having high relative scatter parameter values. The middle row show Blazhko candidates with low relative scatter values. The bottom row shows objects with high relative scatter values but not identified as Blazhko candidates using Fourier analysis.}	
\label{fig:LCgrid}
\end{figure}

Example light curves of Blazhko candidate objects, created using the PDM algorithm, are shown in Fig.~\ref{fig:LCgrid}.

\begin{figure}[ht]
  \centering
   	\resizebox{\hsize}{!}{\includegraphics{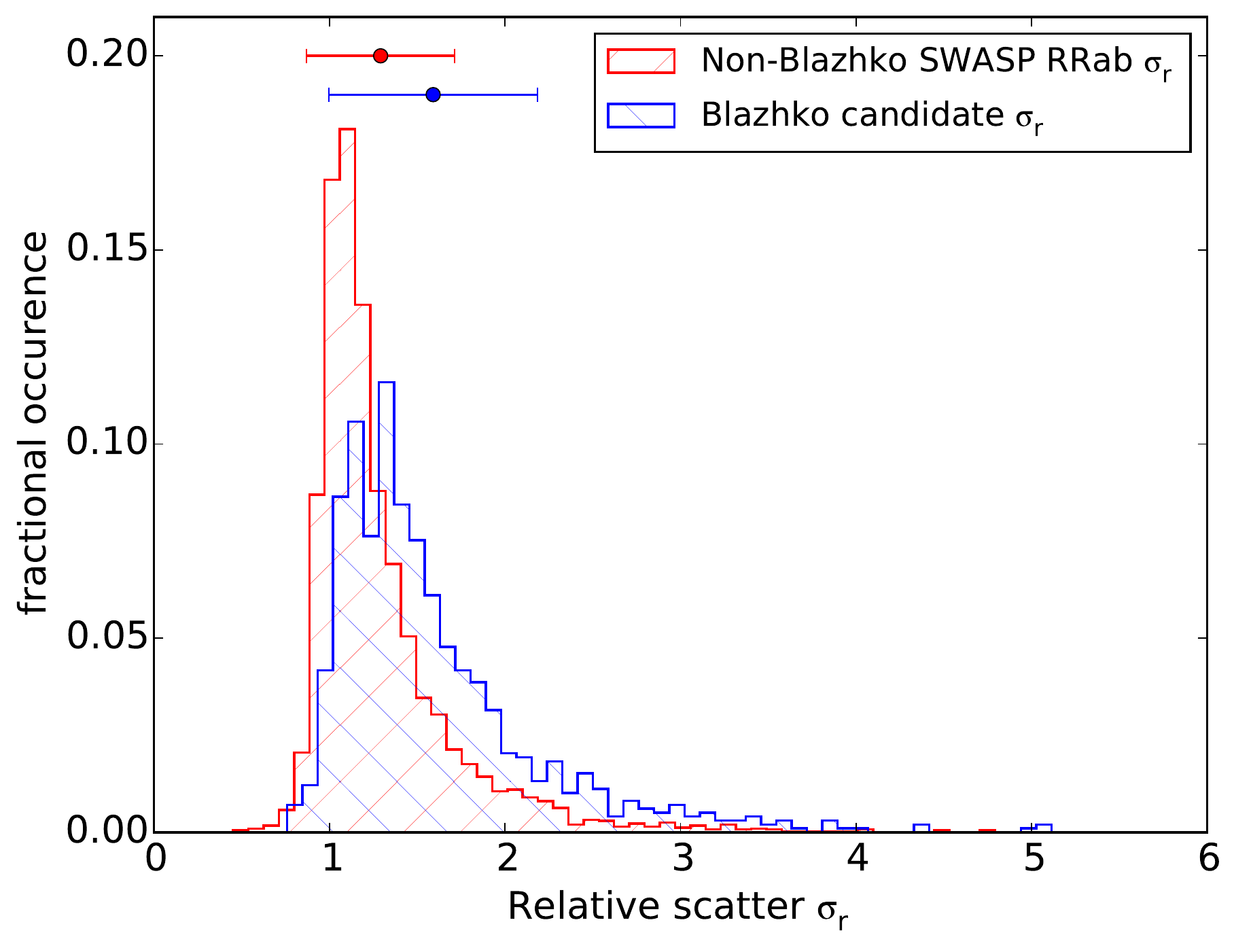}}
	\caption{Distribution of relative scatter $\sigma_r$ for Blazhko and non-Blazhko populations.The error bars represent the mean and standard deviation for each population.}
   \label{fig:relativescattercomparison}
\end{figure}

After analysis of Blazhko and non-Blazhko subsets, the suspected relation between the relative scatter and the Blazhko effect in the time domain is inconclusive. Fig.~\ref{fig:relativescattercomparison} shows that despite a higher number of Blazhko objects having relatively high levels of scatter at the peak, they have very similar lower limits. The non-Blazhko subset has a mean relative scatter of 1.29 and uncertainty of 0.42, and the Blazhko candidates have a mean relative scatter of 1.59 with an uncertainty of 0.59. The lack of bimodality makes it impossible to define a threshold in the relative scatter above which Blazhko candidates can be reliably identified.

\begin{figure}[h]
\centering
  	 \resizebox{\hsize}{!}{\includegraphics{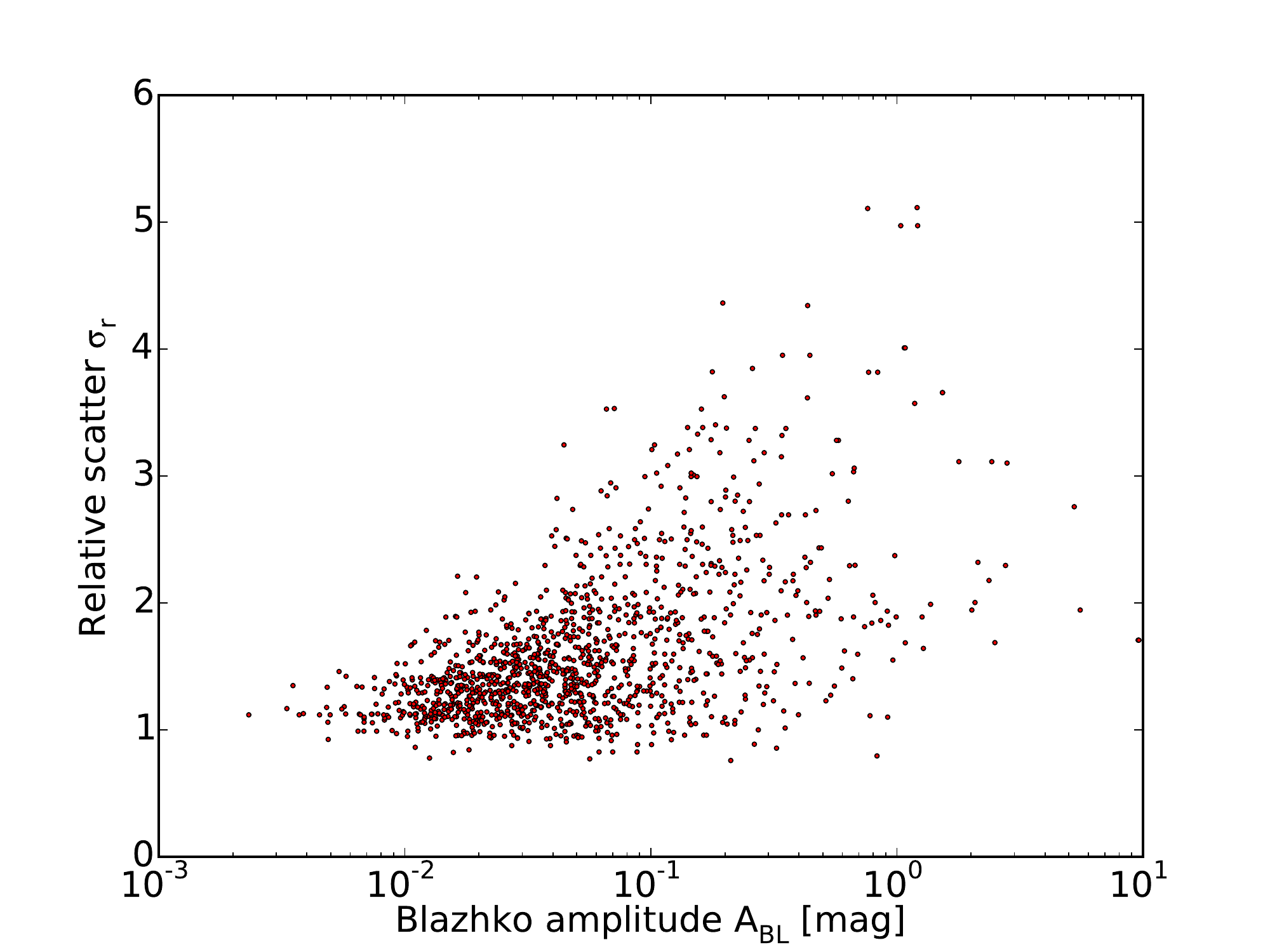}}
      \caption{Comparison of the relative scatter parameter from the phase folded light curves and the Blazhko amplitude from the {\sc CLEAN} power spectra. Despite a visible lack of correlation this relation has a Spearman value of 0.5.}
\label{fig:relSDvsA_BL}
\end{figure}

Fig.~\ref{fig:relSDvsA_BL} shows an increasing spread of relative scatter values with increasing Blazhko power. This produces a moderate monotonic correlation with a Spearman $\rho$ coefficient of 0.50 (p-value $=0.0$)

\begin{figure*}
\centering
	\subfloat[][Non-Blazhko candidates]{\includegraphics[width=7cm]{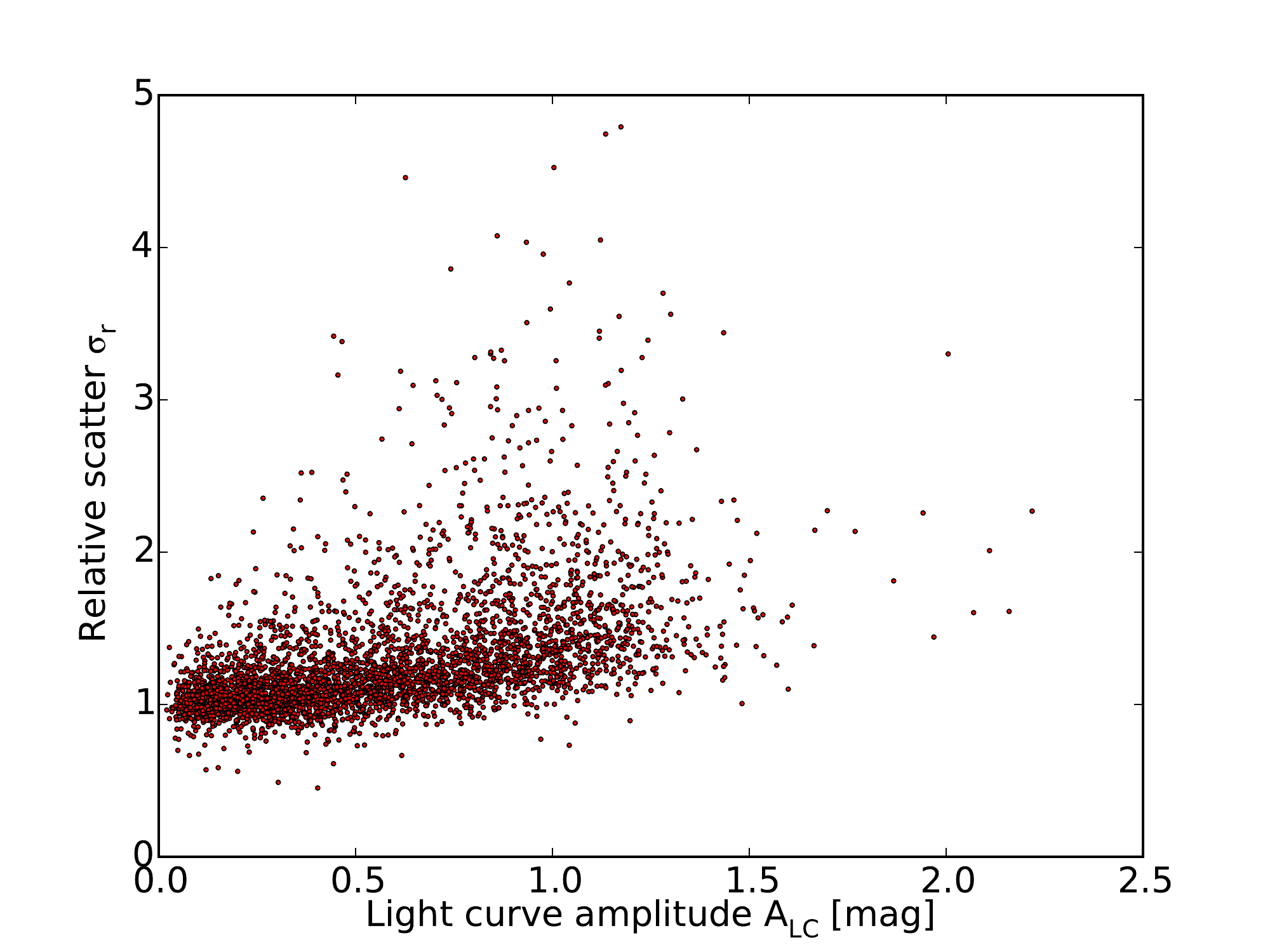}\label{fig:nonBlazRelSDvsA_LC}}
     \subfloat[][Blazhko candidates]{\includegraphics[width=7cm]{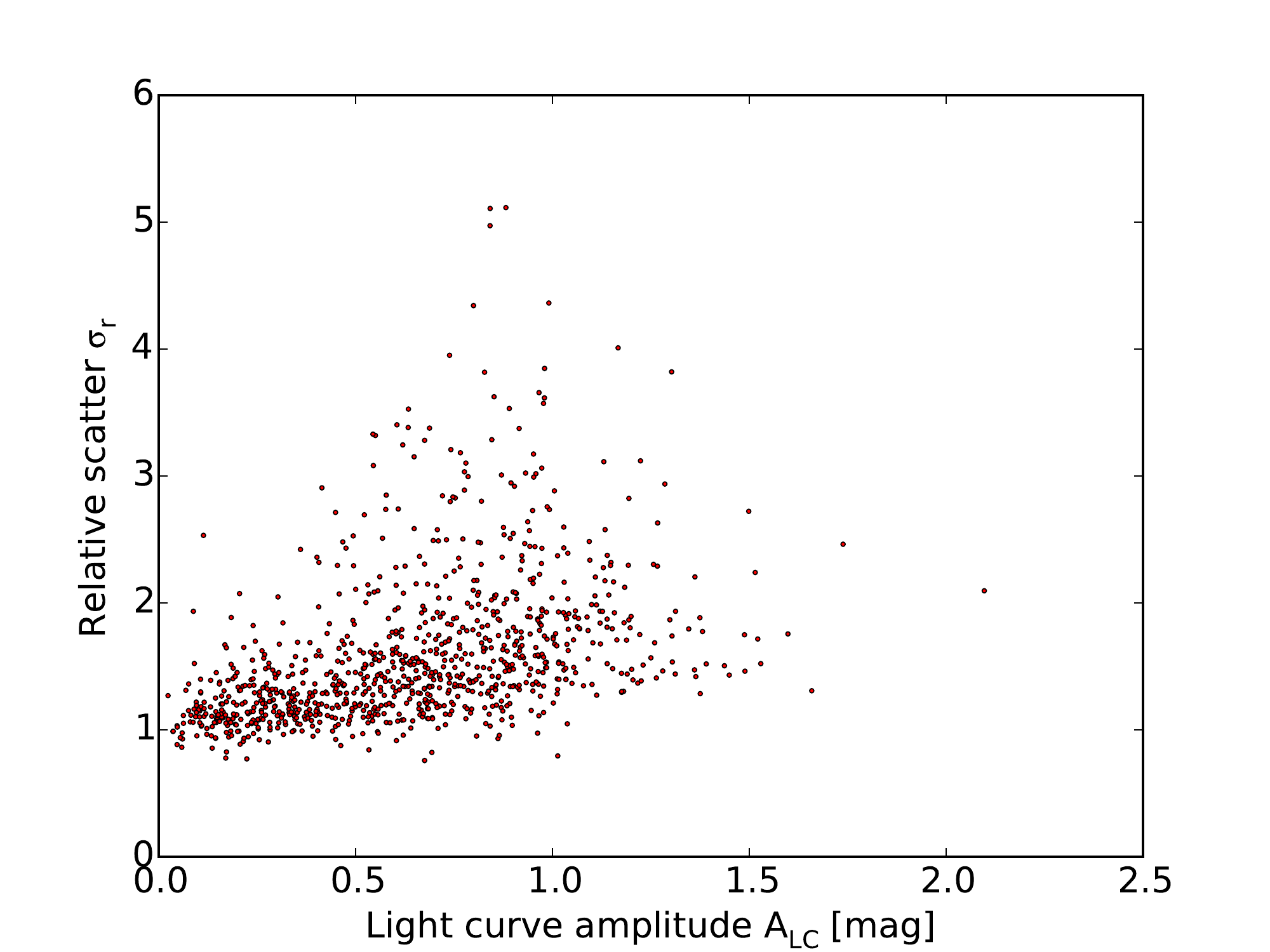}\label{fig:BlazRelSDvsA_LC}}
 	 \caption{Comparison of relative scatter against pulsation period for non-Blazhko and Blazhko populations showing a slight dependence on pulsation amplitude for the relative scatter parameter at high amplitudes.}
 	\label{fig:relSDvsA_LC}
\end{figure*}

The increase in the spread of relative scatter with pulsation amplitude for both non-Blazhko (Fig.~\ref{fig:nonBlazRelSDvsA_LC}) and Blazhko objects (Fig.~\ref{fig:BlazRelSDvsA_LC}) is reflected in their Spearman coefficients of $0.61$ (p-value $=0.0$) and $0.56$ (p-value $=0.0$) respectively. This similarity between populations implies that the amount of scatter at the peak of RRab light curves is dependent only on pulsation amplitude, and is not a signature characteristic of the Blazhko effect.

\begin{figure*}
\centering
	\subfloat[][Blazhko period]{\includegraphics[width=7cm]{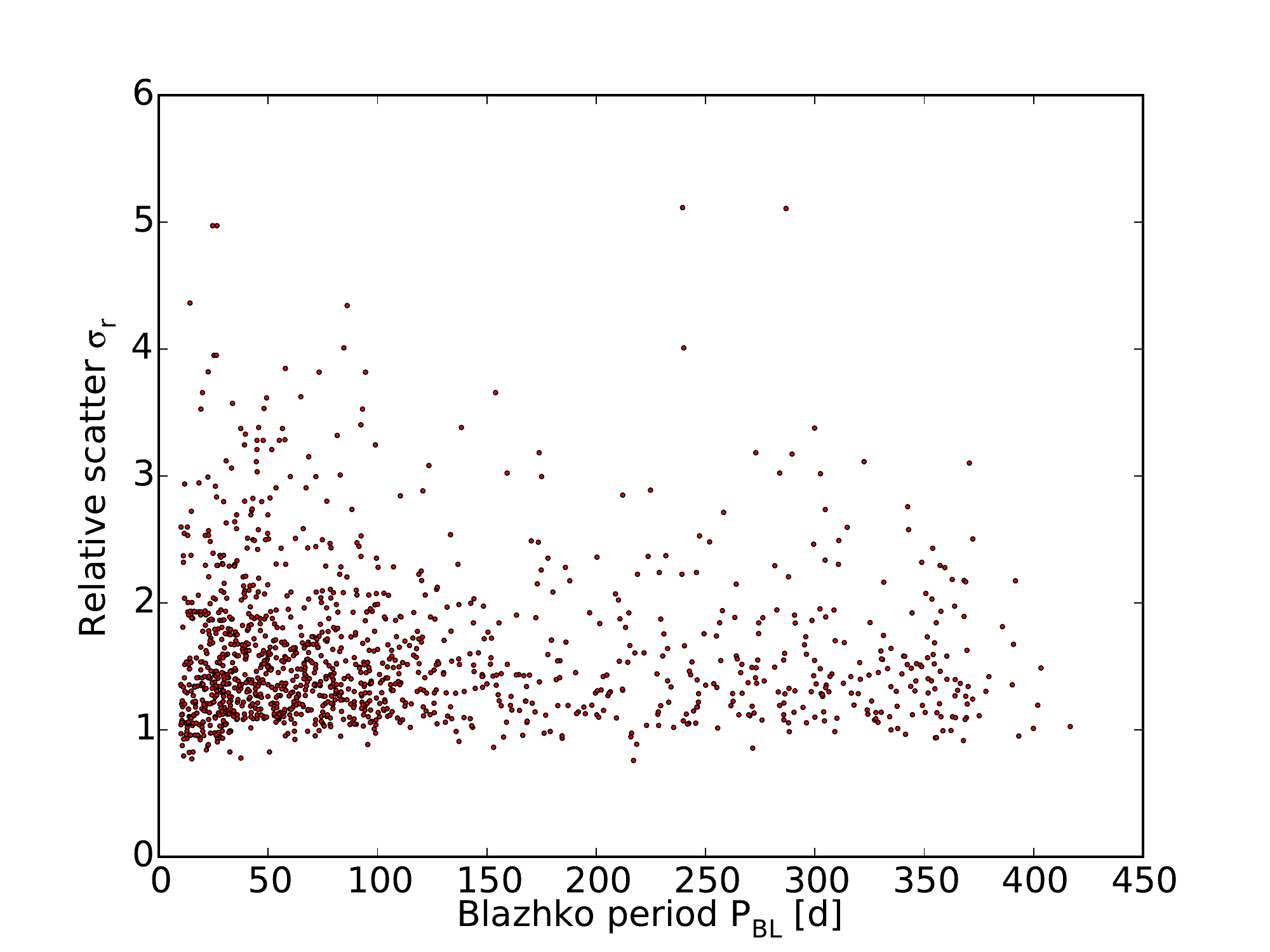}\label{fig:relSDvsP_BL}}
     \subfloat[][Pulsation period]{\includegraphics[width=7cm]{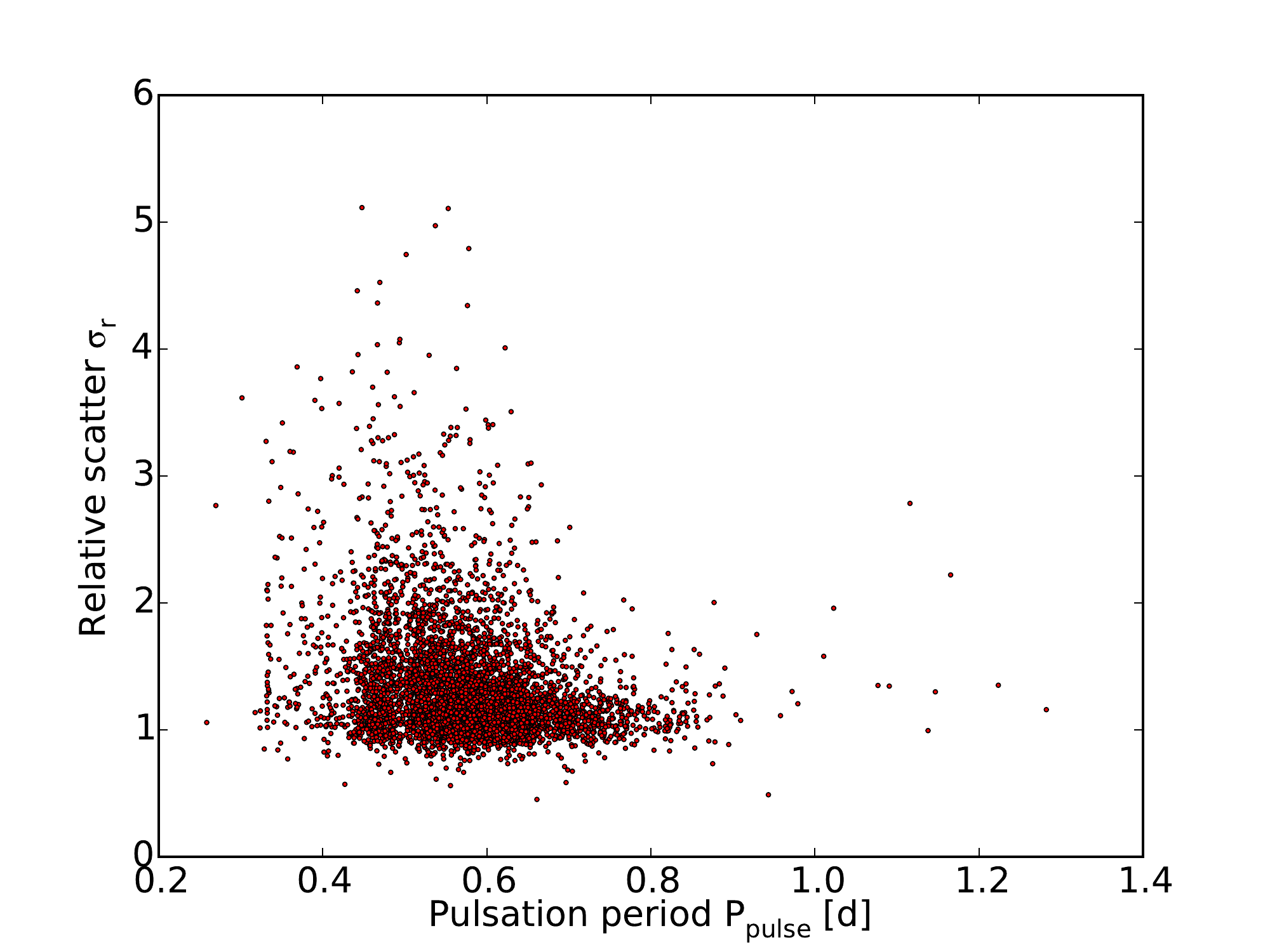}\label{fig:relSDvsP_pulse}}
 	 \caption{Relative scatter parameter against Blazhko and pulsation periods. There appears to be no correlation between the level of scatter at the light curve peak, $\sigma_r$, and the Blazhko period $\rm{P_{BL}}$. There is also no correlation of this scatter with the pulsation period with most objects having a low relative scatter parameter and pulsation periods typical of RRab stars.}
 	\label{fig:relSDvsA_Pulse}
\end{figure*}

\cite{benko_connection_2014} refer to a possible link between the Blazhko period and the amplitude of the AM effect, the so-called "envelope function". The relative scatter parameter described in Sec.~\ref{sec:TimeDomain} is our equivalent measure of this envelope function. However, we are unable to confirm the correlation between the Blazhko period and AM in this study as no correlation can be seen in Fig.~\ref{fig:relSDvsP_BL} and the Spearman coefficient is 0.02 (p-value $= 0.56$). Likewise, there is no correlation between the relative scatter parameter and the main pulsation period as shown in Fig.~\ref{fig:relSDvsP_pulse} where there appears to be a wide range of scatter values around the typical RRab pulsation period of roughly half a day. This relation has a Spearman coefficient of $-0.24$	and p-value of 0.

\subsection{Time domain examples of strong AM}
\label{sec:PDMexamples}

The bottom row of Fig.~\ref{fig:LCgrid} shows examples of objects where strong AM produces several distinct stages in the changing shape of the folded light curve over time. A visual search was made on the catalogue of 4963 SWASP RR Lyrae light curves for more examples. We found 36 light curves that showed this level of AM with relatively little or no phase modulation. In some cases SWASP observations allow the AM of Blazhko candidates' folded light curves to be inspected over the course of an individual cycle.

\begin{figure}[ht]
	\def\tabularxcolumn#1{m{#1}}
	\begin{tabularx}{\linewidth}{@{}cXX@{}}
	\begin{tabular}{ccc}
\subfloat[][2006]{\includegraphics[width=0.32\hsize]{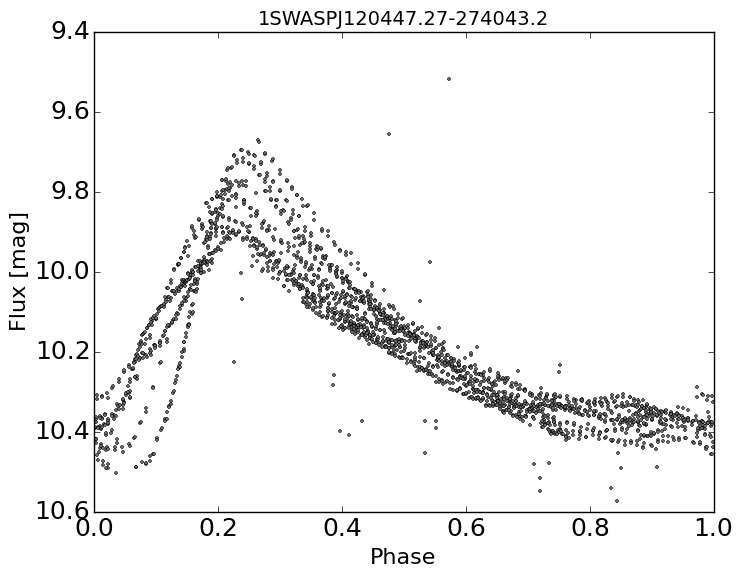} }
    \subfloat[][2007]{\includegraphics[width=0.32\hsize]{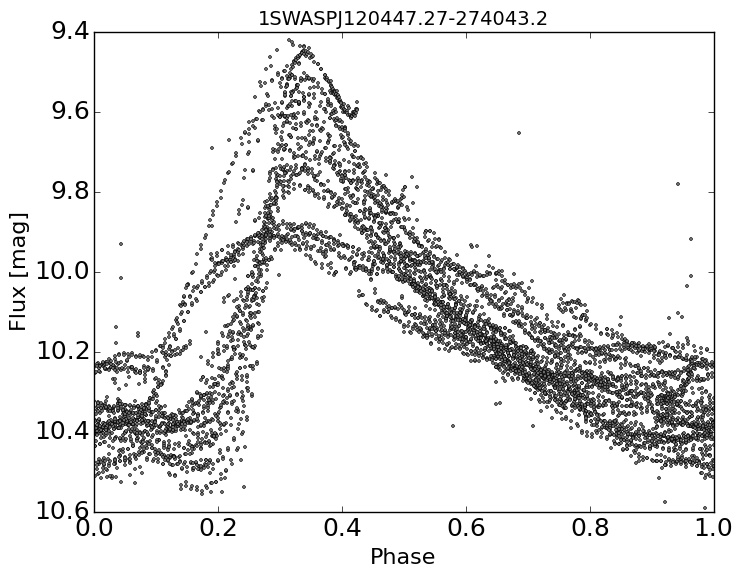} }
    \subfloat[][2008]{\includegraphics[width=0.32\hsize]{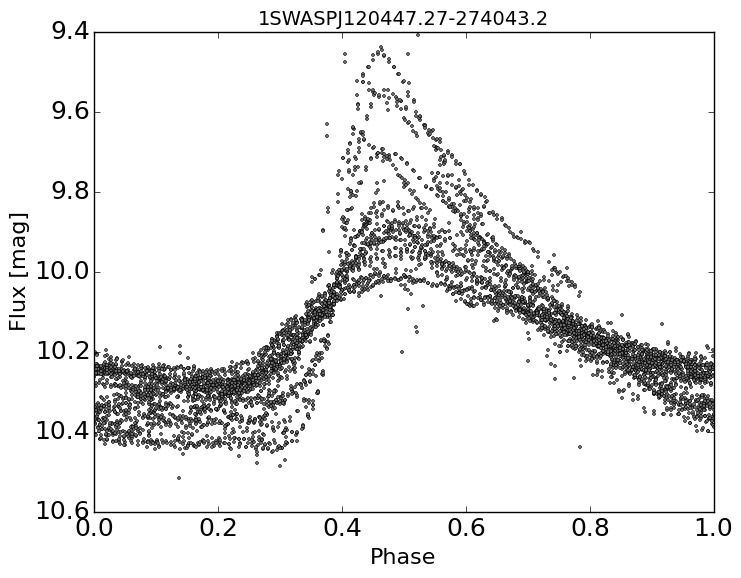} }\\
\subfloat[][2009]{\includegraphics[width=0.32\hsize]{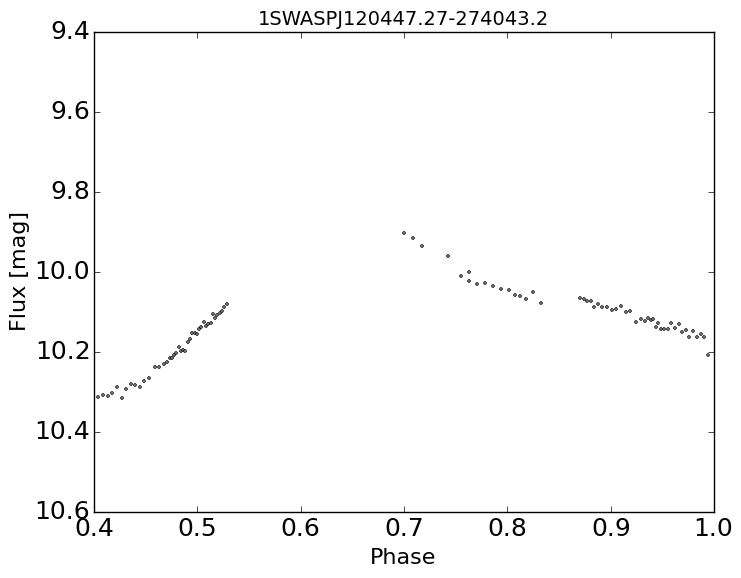} }
   \subfloat[][2010]{\includegraphics[width=0.32\hsize]{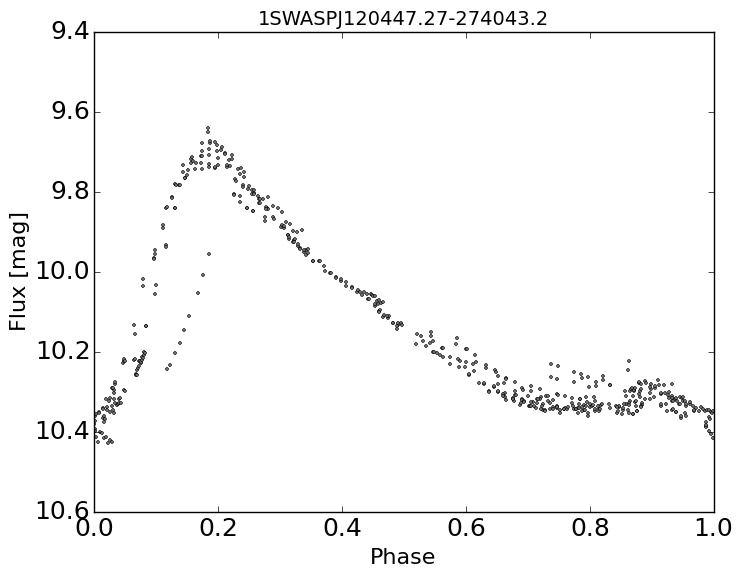} }
  \subfloat[][2011]{\includegraphics[width=0.32\hsize]{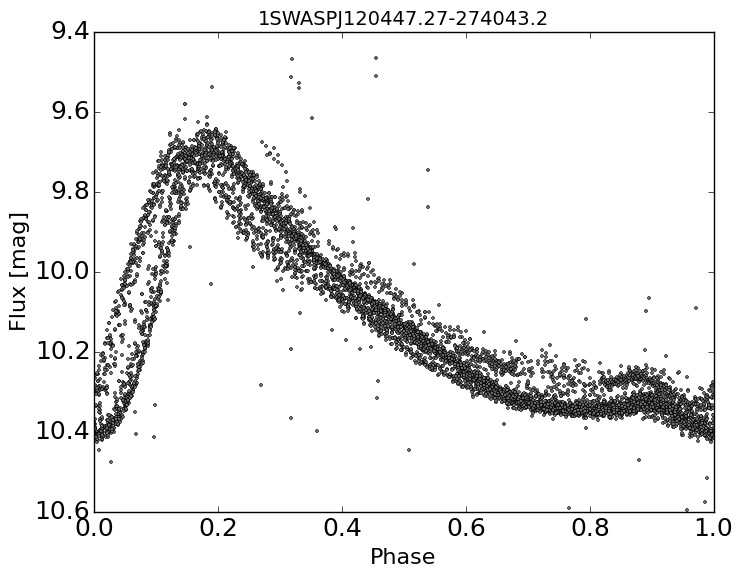} } \\
 \subfloat[][2012]{\includegraphics[width=0.32\hsize]{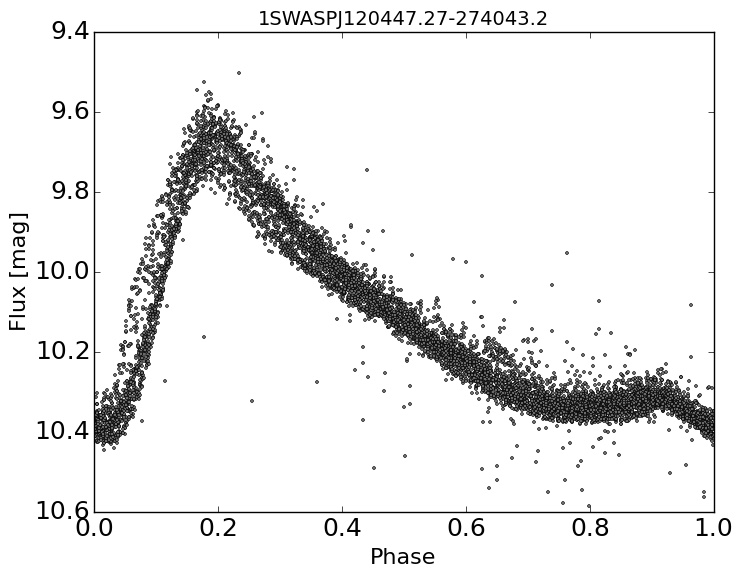} }
	\end{tabular}
	\end{tabularx}
\caption{Yearly light curves of \object{1SWASPJ120447.27-274043.2} from 2006 to 2012 showing how the amplitude of the light curve decreased from 2006 to 2008 and increased again to 2010. Flux levels are scaled in each case to the peak flux.}	
\label{fig:mode_sw}
\end{figure}

\begin{figure}[ht]
	\def\tabularxcolumn#1{m{#1}}
	\begin{tabularx}{\linewidth}{@{}cXX@{}}
	\begin{tabular}{ccc}
  \subfloat[][209]{\includegraphics[width=0.32\hsize]{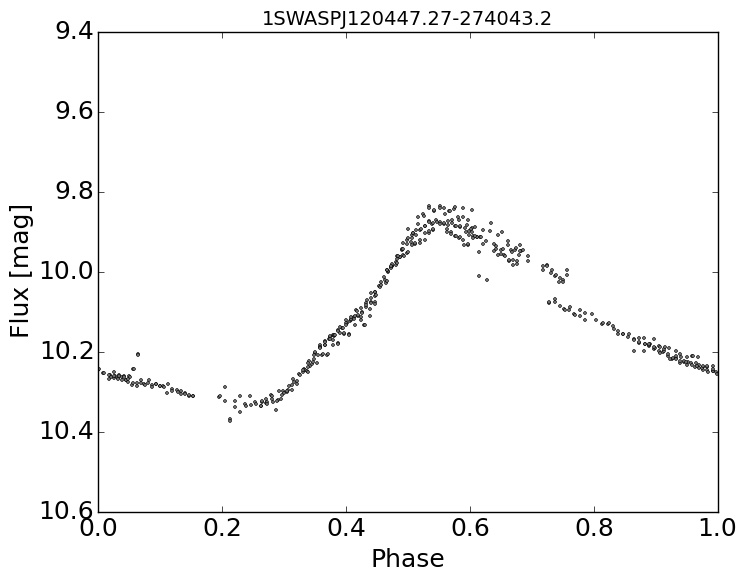} }
  \subfloat[][211]{\includegraphics[width=0.32\hsize]{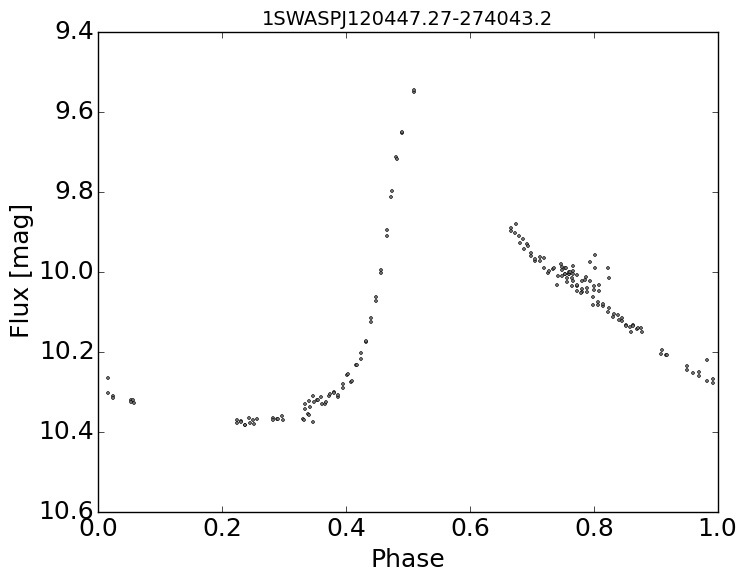} }
  \subfloat[][212]{\includegraphics[width=0.32\hsize]{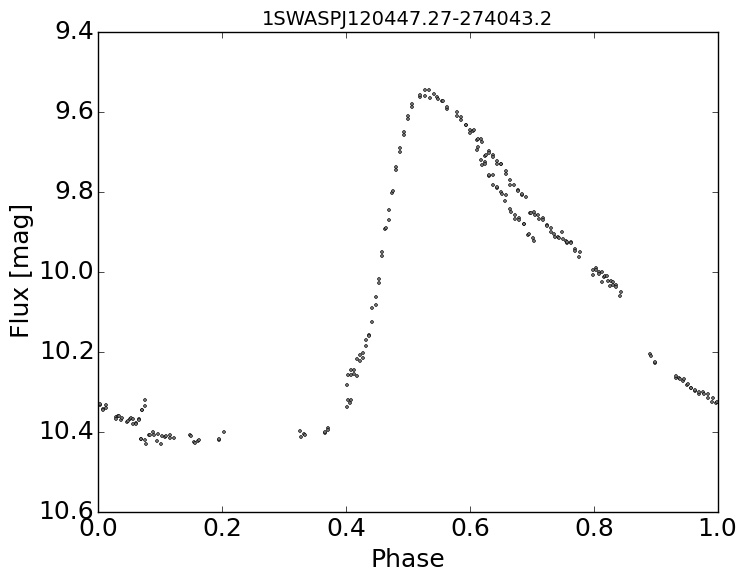} }\\
  \subfloat[][213]{\includegraphics[width=0.32\hsize]{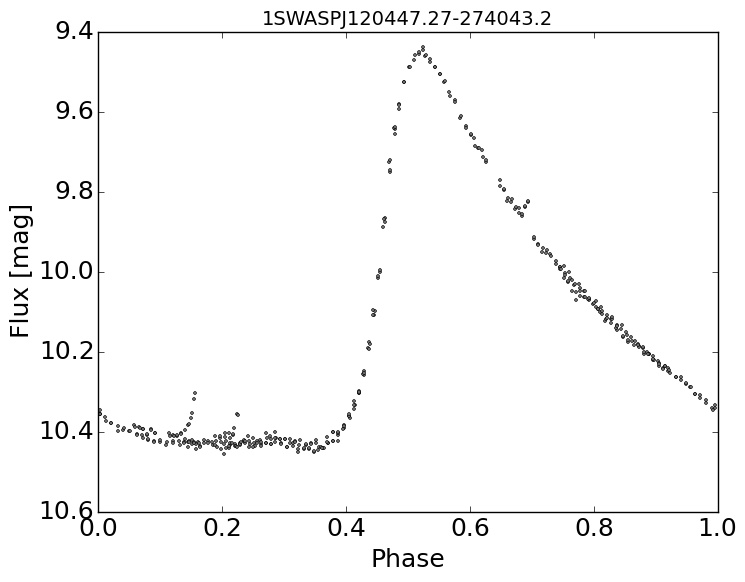} }
  \subfloat[][215]{\includegraphics[width=0.32\hsize]{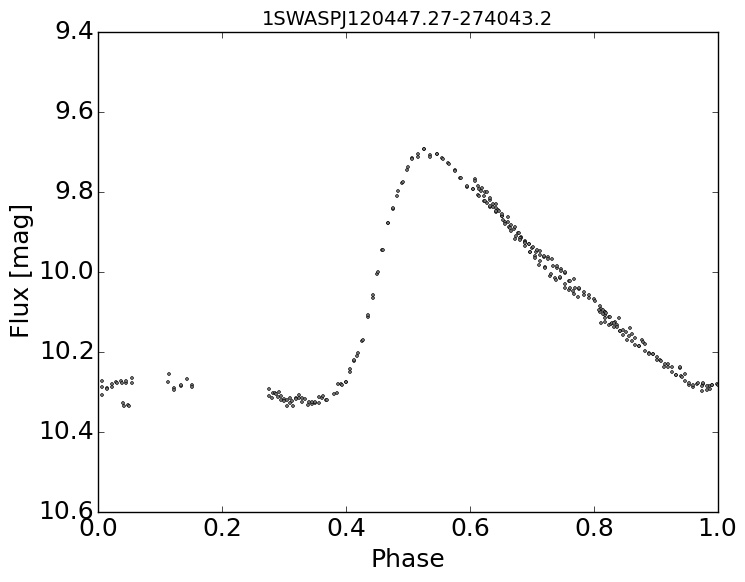} }
  \subfloat[][216]{\includegraphics[width=0.32\hsize]{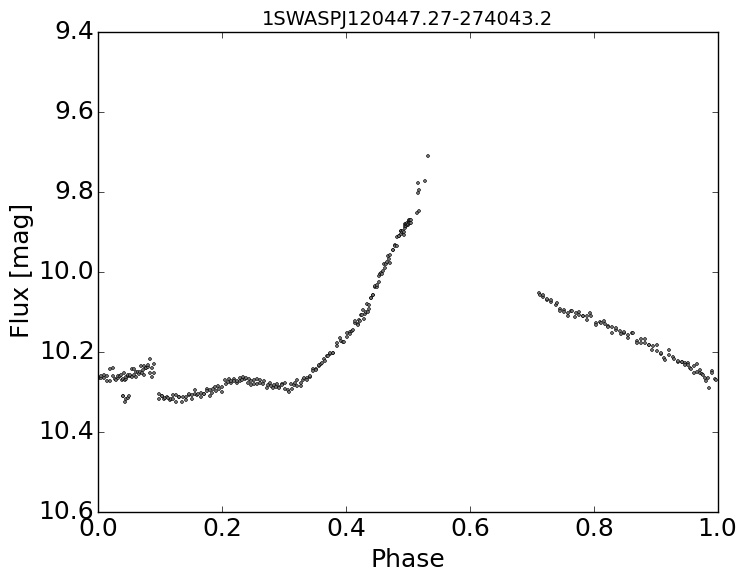} } \\
  \subfloat[][217]{\includegraphics[width=0.32\hsize]{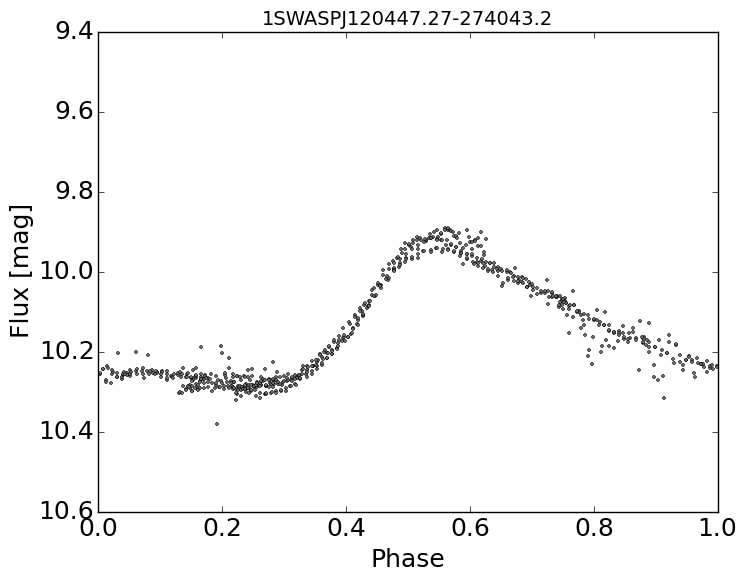} }
  \subfloat[][218]{\includegraphics[width=0.32\hsize]{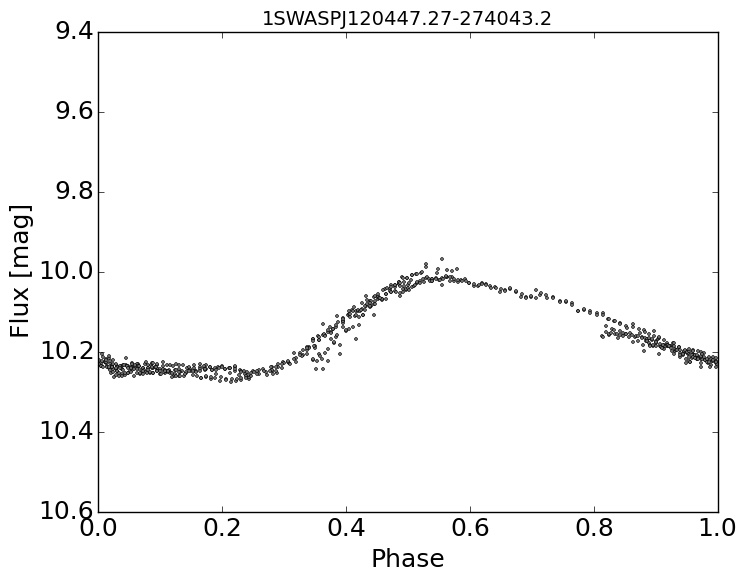} }
  \subfloat[][219]{\includegraphics[width=0.32\hsize]{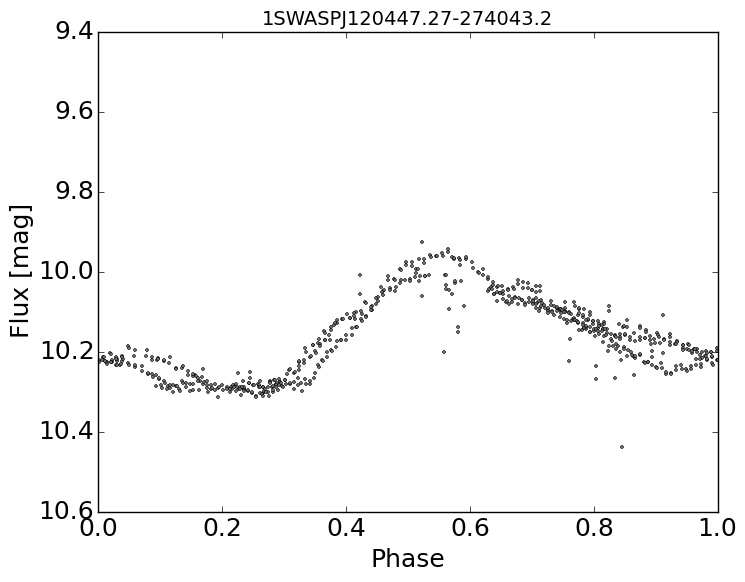} } \\
  \subfloat[][220]{\includegraphics[width=0.32\hsize]{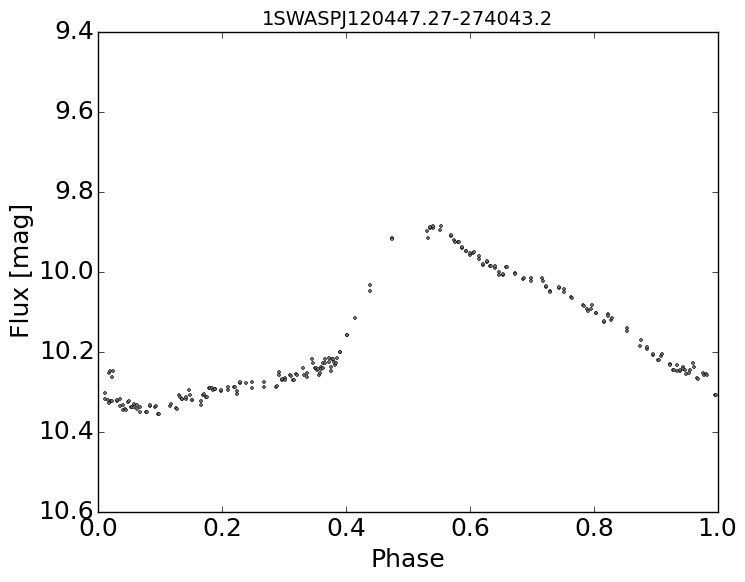} }
  \subfloat[][221]{\includegraphics[width=0.32\hsize]{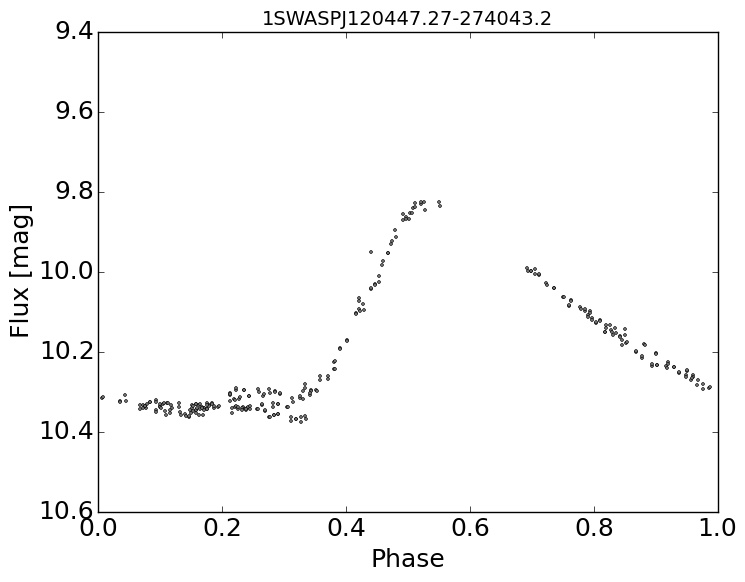} }
  \subfloat[][223]{\includegraphics[width=0.32\hsize]{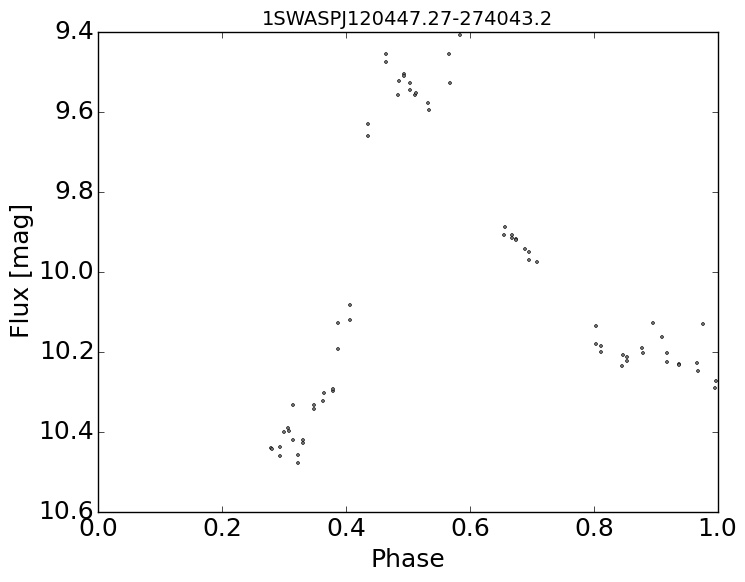} } \\
  \subfloat[][225]{\includegraphics[width=0.32\hsize]{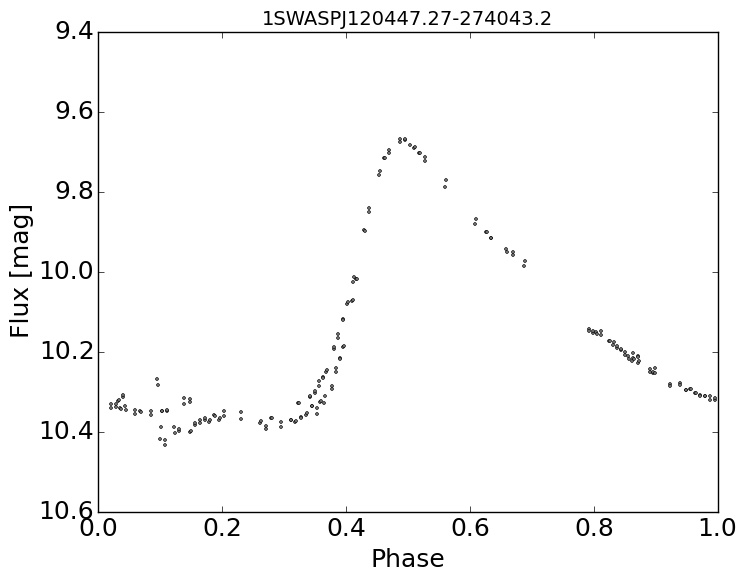} }
	\end{tabular}
	\end{tabularx}
\caption{Weekly sections from 2008 of the light curve of \object{1SWASPJ120447.27-274043.2} for the weeks with enough data for light curve plotting. The week number from the start of the light curve is given under each plot and flux is in magnitudes. All weekly sections of the light curve are folded on the same pulsation period of 0.65029929 days with no change in the phase offset.}
\label{fig:weeklyAM}
\end{figure}

The observations for the 36 extreme AM candidates were split into their constituent years to give each season of observation, as shown in Fig.~\ref{fig:mode_sw} for \object{1SWASPJ120447.27-274043.2}. In all cases the seasonal light curves were phase folded at the same average period found for that object using the PDM program described in subsection \ref{sec:PhaseFoldedLC}. The pulsation periods of the 36 candidates range from 0.347 to 0.668~d. The resultant well--phased light curves means that there was little, if any, period change between years (ruling out any form of mode--switching). \object{1SWASPJ120447.27-274043.2} still showed multiple pulsation profiles overlayed for individual years so this light curve was split into weekly sections, as shown in Fig.~\ref{fig:weeklyAM}, in order to observe how rapidly its light curve was changing shape. It may be seen that significant changes in amplitude and shape of the pulsation occur on timescales as short as a week in this case.

\section{Investigations of parameter correlations}
\label{sec:analysis}
Several of the basic characteristics of the SuperWASP RRab and their Blazhko effect parameters were compared to search for any correlations.

\begin{figure*}
\centering
	\subfloat[][Pulsation periods]{\includegraphics[width=7cm]{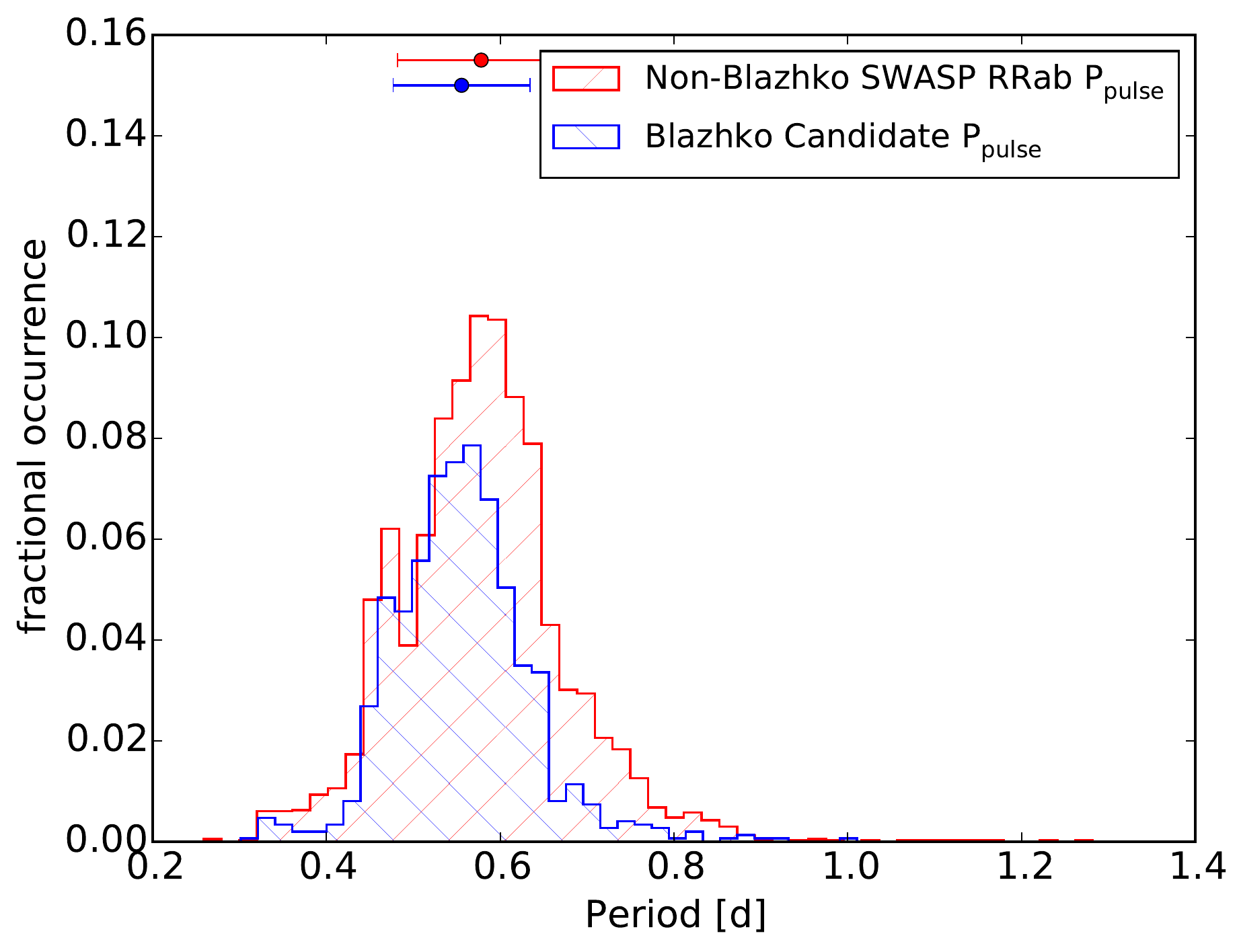}\label{fig:pulseperiods}}
     \subfloat[][Pulsation amplitudes]{\includegraphics[width=7cm]{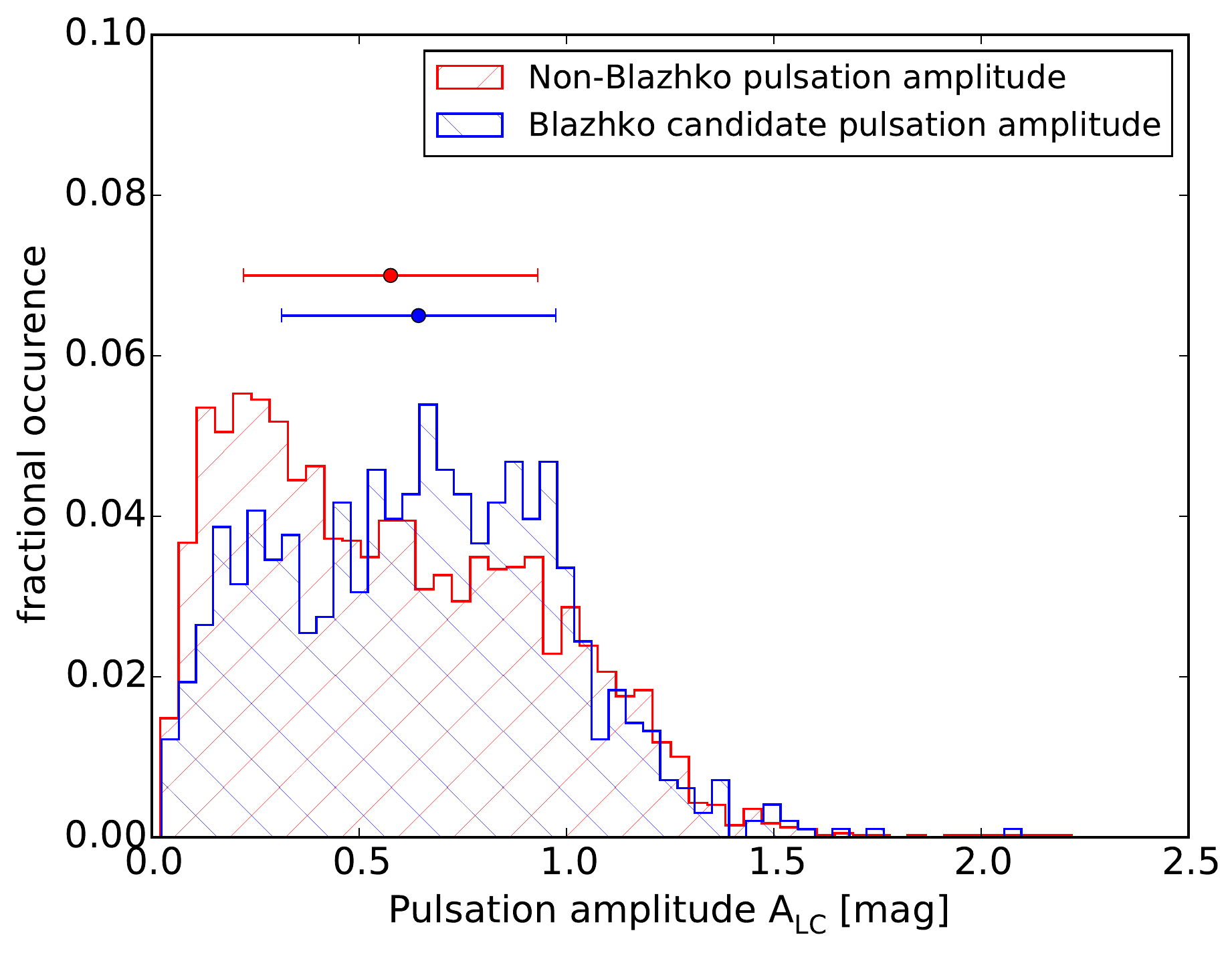}\label{fig:pulseampcomparison}}
 	 \caption{Comparison of pulsation periods and amplitudes for SWASP Blazhko and non-Blazhko objects. The mean and $1\sigma$ deviation are shown by the horizontal error bars, where the colours correspond to the respective populations.} 
 	\label{fig:pulsecomparison}
\end{figure*}

The pulsation periods for SWASP RRab and Blazhko objects are compared in Fig.~\ref{fig:pulseperiods}. The slight deficit of pulsation periods between 0.45 and 0.50~d is thought to be an effect of the ground--based sampling regime. The mean pulsation period of the Blazhko candidate subset is 0.56~d with a standard deviation of 0.08~d. The non-Blazhko subset's mean pulsation period is 0.58~d, with a standard deviation of 0.10~d. The Blazhko candidates' pulsation periods range from 0.30 to 1.01~d, whereas the non-Blazhko subset have a wider range from 0.26 to 1.28~d.

Fig.~\ref{fig:pulseampcomparison} shows that Blazhko and non-Blazhko effect objects have similar ranges of pulsation amplitudes, apart from a very small number of high pulsation amplitude non-Blazhko objects. The slight deficit of pulsation amplitudes between 0.2 and 0.5 mag in the Blazhko dataset still appears after running a bootstrap algorithm to check if it was a sampling effect. The ability to detect smaller amplitudes suggests that it is not a systematic limitation. The cause of this deficit remains unknown at this stage.

\begin{figure}[ht]
 \centering
   \resizebox{\hsize}{!}{\includegraphics{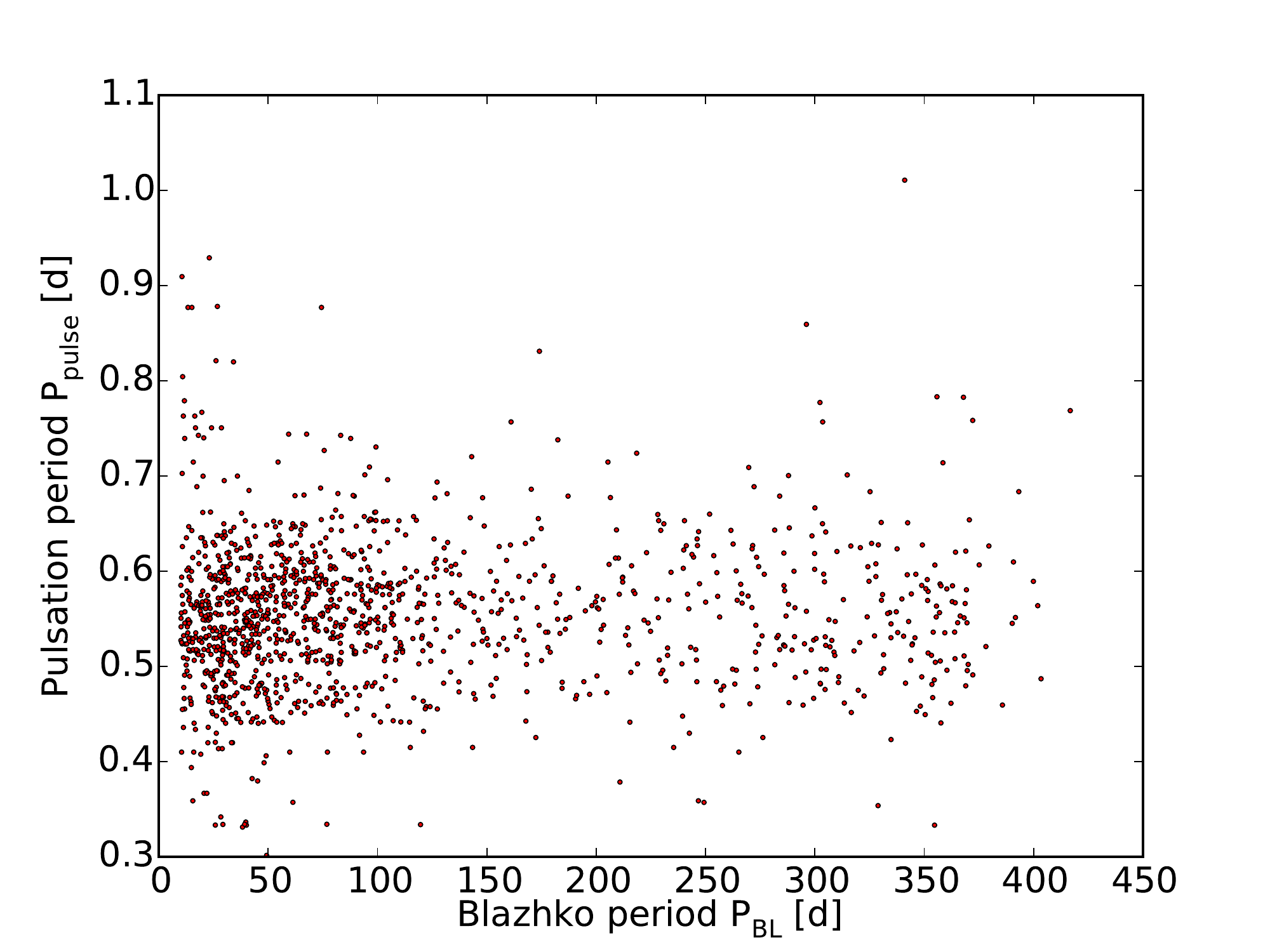}}
      \caption{Comparison of pulsation period ($\rm{P_{pulse}}$) against Blazhko period ($\rm{P_{BL}}$) showing a similar distribution across a wide range of Blazhko periods.}
   \label{fig:P_pulsevsP_BL}
\end{figure}

Blazhko periods were compared to pulsation periods, as shown in Fig.~\ref{fig:P_pulsevsP_BL}, but no correlation was found between these basic characteristics of our Blazhko candidates, with the Spearman rank coefficient of 0.068 (p-value $=0.01$).

\begin{figure*}[ht]
\centering
	\def\tabularxcolumn#1{m{#1}}
	\begin{tabular}{cc}
\subfloat[]{\includegraphics[width=7cm]{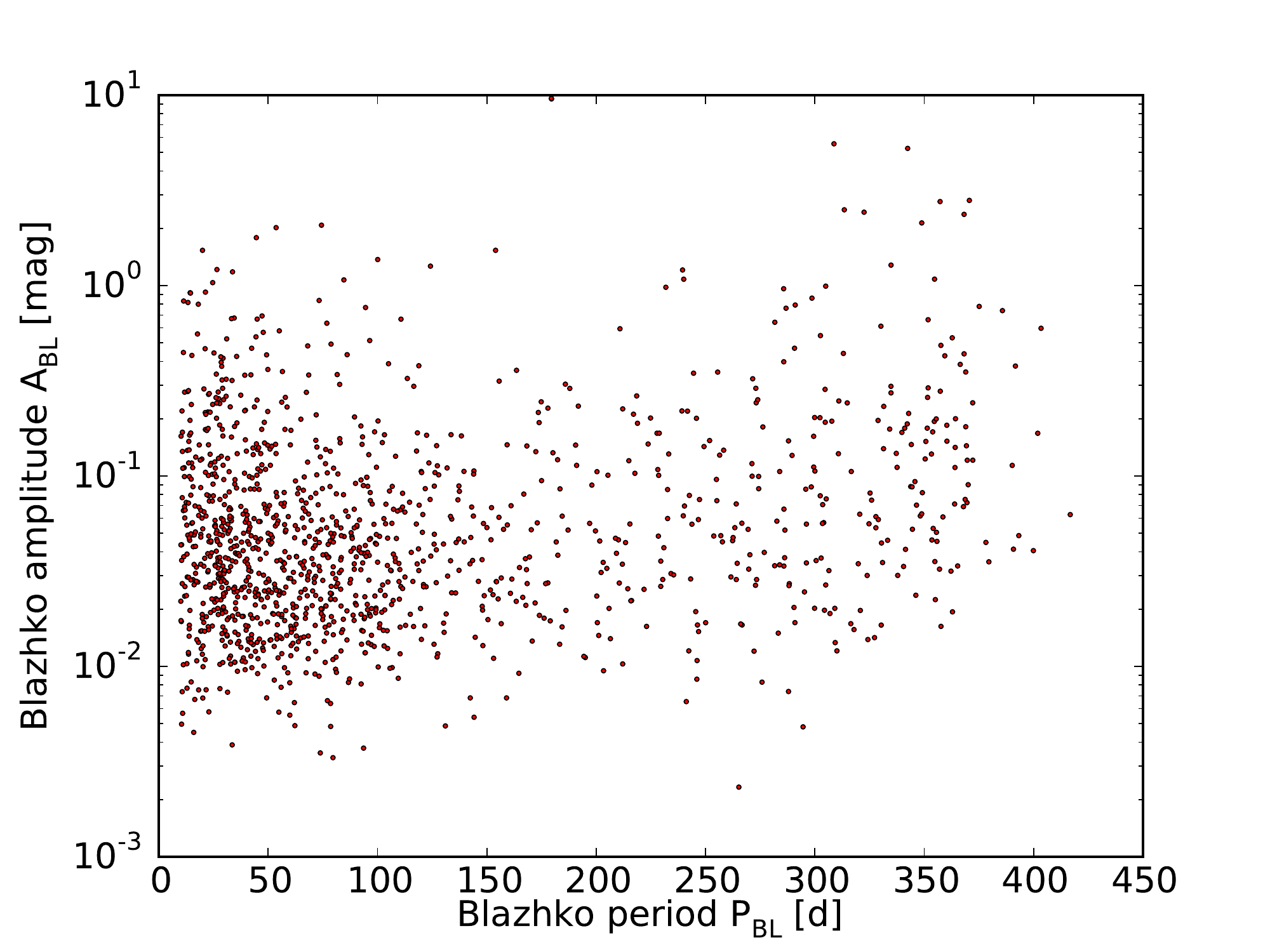}\label{fig:A_BLvsP_BL}}
 \subfloat[]{\includegraphics[width=7cm]{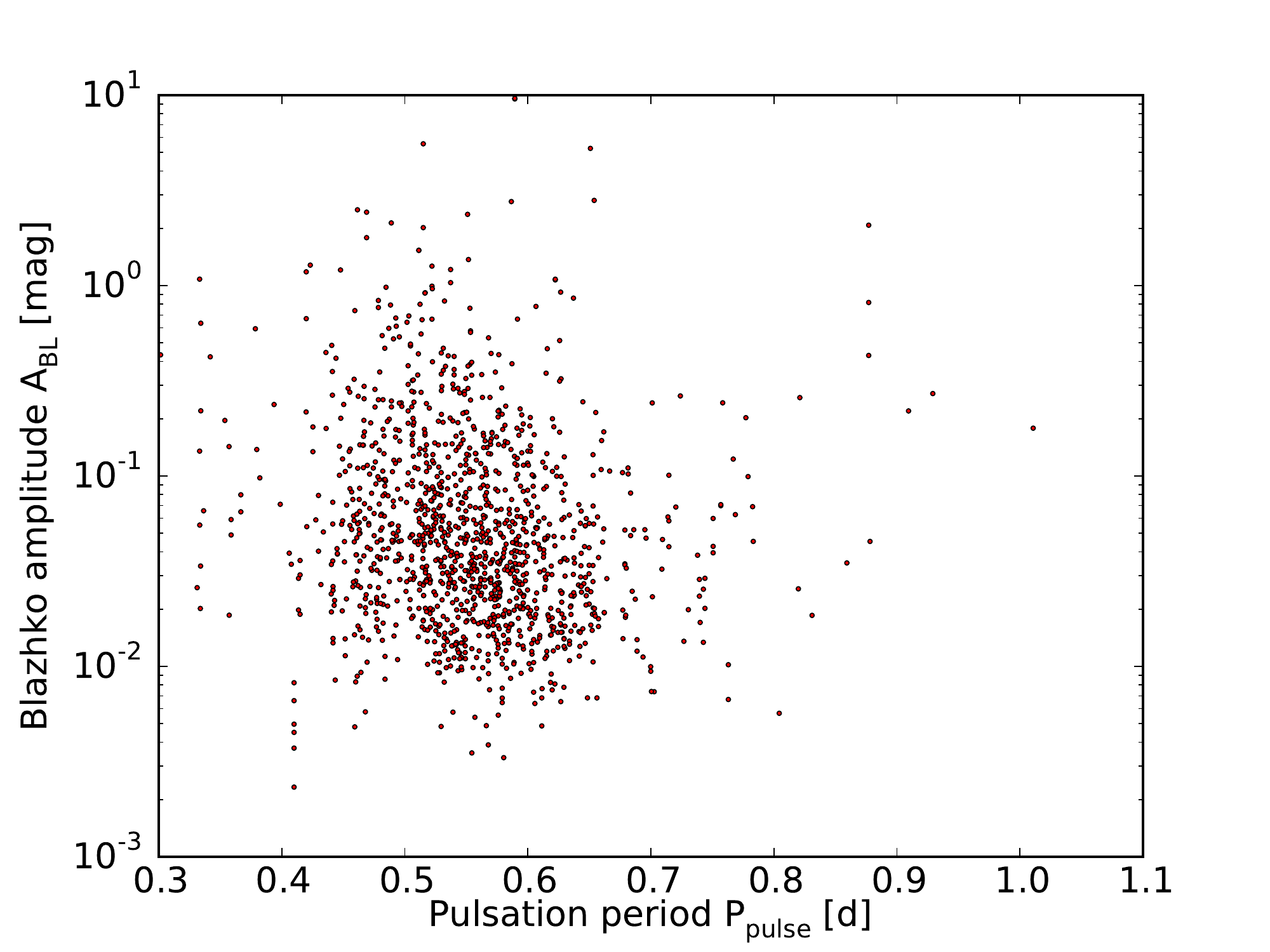}\label{fig:A_BLvsP_pulse}}
    \\
\subfloat[]{\includegraphics[width=7cm]{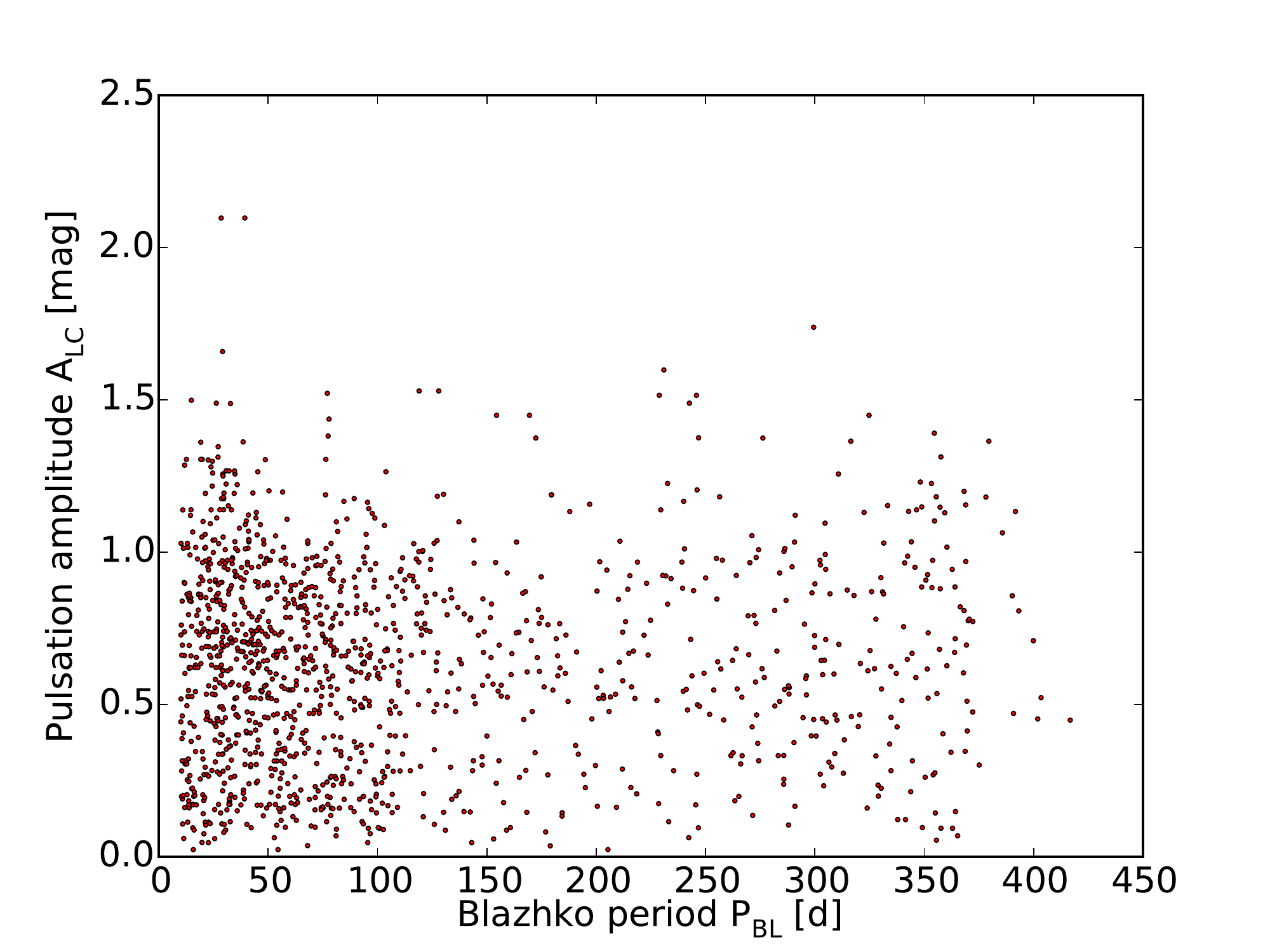}\label{fig:BlazPeriodsvsPulseAmp}}
\subfloat[]{\includegraphics[width=7cm]{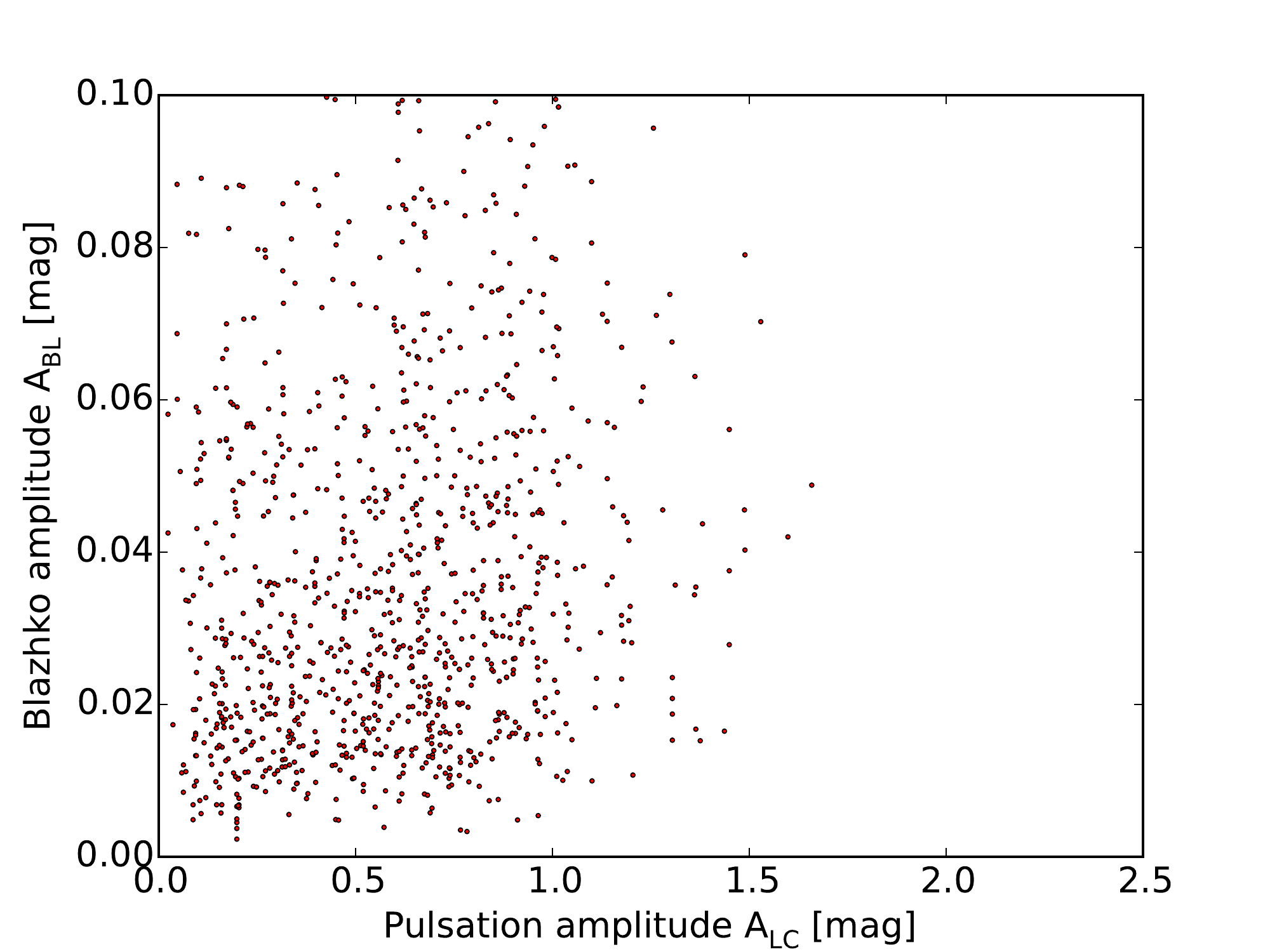}\label{fig:A_pulsevsA_BL}}
	\end{tabular}
\caption{Comparison of the basic light curve parameters of the Blazhko amplitude $\rm{A_{BL}}$, the Blazhko period $\rm{P_{BL}}$, the pulsation period $\rm{P_{pulse}}$. and the pulsation amplitude $\rm{A_{LC}}$}
\end{figure*}

The amplitude of the Blazhko effect, $A_{BL}$, was taken from the amplitude of the fitted modulation frequency returned by Period04 as described in Sec.~\ref{CLEANresults}. This amplitude was compared with the Blazhko period (Fig.~\ref{fig:A_BLvsP_BL}), however, there is a lack of correlation between them over a wide range of values producing a Spearman coefficient of 0.11(p-value $=2\times 10^{-5}$). Additionally, Fig. \ref{fig:A_BLvsP_pulse} shows how the Blazhko amplitude is not related to the main pulsation period either, with a Spearman coefficient of $-0.199$ (p-value $=7.98\times10^{-14}$).

 Finally we compared candidate Blazhko periods and amplitudes with pulsation amplitudes calculated from the peak to peak amplitude of the phase folded light curve, denoted as $A_{LC}$. The pulsation amplitude is shown compared to the Blazhko period in Fig.~\ref{fig:BlazPeriodsvsPulseAmp}, where no correlation is found despite a wide range of candidate  periods. The Spearman value in this case was 0.17 (p-value $=9\times 10^{-4}$). The correlation between the Blazhko amplitude and the pulsation amplitude (Fig.~\ref{fig:A_pulsevsA_BL}) has a strong Spearman coefficient of 0.735 (p-value $=0$). However, {\sc CLEAN}ing a synthetic light curve which included a modulation effect demonstrated that when the main pulsation amplitude was increased, the strength of the modulation peaks in the {\sc CLEAN} power spectrum also increased. Therefore the apparent correlation between Blazhko amplitude and pulsation amplitude may be a systematic effect of the power spectrum and not a property of the Blazhko effect itself.

\section{Discussion}
\label{sec:discussion}
Sec.~\ref{sec:analysis} highlighted the lack of correlation between the basic and modulation parameters of our SWASP RRab population. No relation was found between the pulsation period and the Blazhko period, which concurs with \cite{skarka_bright_2014}, or between the pulsation period and the Blazhko amplitude.
The lower occurrence of longer pulsation periods in the Blazhko candidate subset is in agreement with \cite{prudil_blazhko_2017}; in our case this deficit starts at $\sim{0.6}$~d. However, there is still a large overlap in the range of pulsation period in both Blazhko and non-Blazhko populations.

The 20\% incident rate of Blazhko effects stars in this catalogue is lower than the figures from   ground-based surveys of the Galactic field:
31\% found in \cite{skarka_bright_2014} using a sample of 321 bright stars from ASAS and SuperWASP; and 47\% found by the Konkoly Blazhko Survey \citep{jurcsik_konkoly_2009}. It is also lower than the 39\% occurrence from the space-based Kepler mission \citep{szabo_blazhko_2014}. 
 This may be a consequence of our strict acceptance criteria for recognising a Blazhko candidate, in that we require either a match between a low frequency peak and a sideband peak, or a match between two sideband peaks, in the power spectrum. Alternatively it may be that we are missing objects whose Blazhko signal is very weak and therefore below our noise levels. Our proportion is, however, higher than the 10\% occurrence of the Blazhko effect in first overtone RRc stars in the globular cluster M3 recorded by \cite{jurcsik_modulation_2014}.

Despite matching low frequency peaks to sidebands, some systematic aliases appear to remain at 14 and 29 days judging by the slight increase in the number of Blazhko candidates in these two regions. These are most likely the result of noise occurring in the SuperWASP light curves on the lunar cycle, and as such any candidate Blazhko periods around these values should be treated with caution.

\cite{benko_connection_2014} suggested a link between the period and amplitude of the Blazhko effect but Fig.~\ref{fig:A_BLvsP_BL} shows that this cannot be confirmed in this study. Also, the Blazhko period does not appear to have any correlation with the RRab pulsation amplitude (Fig.~\ref{fig:BlazPeriodsvsPulseAmp}).

The lack of bimodality in the distribution of the relative scatter parameter for all SuperWASP objects (Fig.~\ref{fig:relativescattercomparison}) makes it difficult to produce a threshold value for identifying the Blazhko effect in the time domain without excluding a large number of potential candidates. The increase in the spread of relative scatter with light curve amplitude for both non-Blazhko (Fig.~\ref{fig:nonBlazRelSDvsA_LC}) and Blazhko objects (Fig.~\ref{fig:BlazRelSDvsA_LC}) implies that high relative scatter could be a feature of large RRab pulsations rather than a symptom of the Blazhko effect.
The similarity between plots of relative scatter against Blahzko amplitude and relative scatter against light curve amplitude (Figs.~\ref{fig:relSDvsA_BL} and ~\ref{fig:BlazRelSDvsA_LC}) could merely be due to the strong correlation between Blazhko amplitudes and pulsation amplitudes in power spectra.

The lower section of the folded light curves, used in comparison to the scatter of flux at the peak area when calculating the relative scatter, was affected by modulation more than anticipated. FM in particular increases the horizontal scatter throughout the whole light curve, which in turn raises the level of vertical scatter in the lower comparison area without increasing scatter at the peak, thereby reducing the relative scatter between these two areas. The relative scatter technique therefore only produces high parameter values for objects with high AM but low FM. The large overlap between the Blazhko and non-Blazhko populations suggests that either these AM and FM are not seen in isolation, or there is a lack of a low AM threshold for the Blazhko effect. Despite measuring Blazhko amplitude values as low as just a few mmag, no minimum threshold value to the Blazhko amplitude is reached. This suggests that the Blazhko modulation may be a common but undiscovered effect in many RRab.  More precise observations may detect that Blazhko modulation is a feature of all pulsating stars, but at levels that are too low to detect in SWASP data.

Future work may include isolation of the FM aspect of the Blazhko effect through investigations using O-C techniques.
This could be compared to the relative scatter distribution to investigate the relationship between AM and FM in the Blazhko effect, or to inspect individual cases where only one or other are present.

\section{Conclusions}
\label{sec:conclusions}
A large population of 4963 RRab class RR Lyrae stars were identified from the SuperWASP archive using the initial classifications of \cite{payne_identification_2013}. A bespoke phase dispersion minimisation and epoch folding routine determined pulsation period and amplitude parameters. It also determined a basic measurement of the scatter at the peak relative to elsewhere on the light curve as an attempt at systematically identifying Blazhko effect objects in the time domain. The {\sc CLEAN} routine was used to obtain the power spectra of these objects and 1387 Blazhko candidate periods were discovered for 983 Blazhko candidate objects where there was a match between a low frequency peak and a sideband, or between two sidebands in the power spectrum. Despite (or because of) the large heterogeneous sample of objects, no clear correlations between parameters were identified. The lower rate of occurrence of the Blazhko effect in long pulsation period objects was corroborated in our study.

\textit{Note in Press}:
The 3397 ‘previously unknown’ RR Lyrae stars presented here are those not currently included in the GCVS; some of these may additionally exist in other lists such as the AAVSO International Variable Star Index. Since finalising this paper, we have also been alerted to the fact that a more complete CSS catalogue of 14,362 RR Lyrae stars exists (Drake, et al. 2013, ApJ, 763, 32 and Drake, et al. 2013, ApJ, 765, 154). A further 1866 of our 4963 SuperWASP RR Lyrae appear in this larger CSS catalogue.

\begin{acknowledgements}
The authors thank the referee for the many useful comments and suggestions which helped to improve this work. 
The WASP project is currently funded and operated by Warwick University and Keele University, and was originally set up by Queen's University Belfast, the Universities of Keele, St. Andrews and Leicester, the Open University, the Isaac Newton Group, the Instituto de Astrofisica de Canarias, the South African Astronomical Observatory and by STFC.  This work has made use of the Vizier database, operated at CDS, Strasbourg, France, the cross-match service provided by CDS, Strasbourg, and M. Skarka's BlaSGalF online list of Blazhko effect stars: (\url{http://www.physics.muni.cz/~blasgalf/}. 
\end{acknowledgements}

\bibliography{BlazhkoCatalogue170624}

\bibliographystyle{aa} 

\clearpage

\end{document}